\documentclass[journal]{IEEEtran}
\usepackage{cite}
\usepackage{amsfonts, amssymb, amsthm, bm, bbm, graphicx, relsize, multirow, booktabs, tikz,subfigure,soul}
\usepackage{ucs}
\usepackage[cmex10]{amsmath,mathtools}
\usepackage[american]{babel}
\usepackage[T1]{fontenc}
\usepackage{tabulary}
\usepackage{algorithmic, algorithm}
\usepackage[multiple]{footmisc}

\newcommand{\beq}{\begin{equation}}
	\newcommand{\eeq}{\end{equation}}
\newcommand{\bH}{\mathbf{H}}
\newcommand{\bA}{\mathbf{A}}
\newcommand{\bD}{\mathbf{D}}
\newcommand{\bW}{\mathbf{W}}
\newcommand{\bw}{\mathbf{w}}
\newcommand{\bPhi}{\mathbf{\Phi}}
\newcommand{\bY}{\mathbf{Y}}
\newcommand{\by}{\mathbf{y}}
\newcommand{\bh}{\mathbf{h}}
\newcommand{\bp}{\mathbf{p}}

\newcommand{\ds}{\displaystyle}
\newcommand{\norm}[1]{\left\lVert#1\right\rVert}    

\usepackage{graphicx}

\begin{document}

This paper has been accepted for publication on the IEEE Transactions on Cognitive Communications and Networking, special issue on "Intelligent Surfaces for Smart Wireless Communications". It was originally submitted for publication on September 15, 2020, and revised on February 3, 2021 and on March 12, 2021. It has been finally accepted for publication on March 14, 2021. 

\bigskip

\copyright 2021 IEEE. Personal use of this material is permitted. Permission from IEEE must be obtained for all other uses, in any current or future media, including reprinting/republishing this material for advertising or promotional purposes, creating new collective works, for resale or redistribution to servers or lists, or reuse of any copyrighted  component of this work in other works.

\newpage

	\bstctlcite{IEEE_nodash:BSTcontrol}
	\title{RIS Configuration, Beamformer Design, and Power Control in Single-Cell and Multi-Cell Wireless Networks}
	%
	%
	%
	
	\author{Stefano Buzzi,~\IEEEmembership{ Senior Member,~IEEE,}
		Carmen D'Andrea,~\IEEEmembership{ Member,~IEEE,} Alessio Zappone,~\IEEEmembership{ Senior Member,~IEEE,} Maria Fresia, 
		Yong-Ping Zhang, and Shulan Feng
		\thanks{This paper has been partly presented at 2020 IEEE PIMR C.}    \thanks{This work was supported by HiSilicon through cooperation agreement YBN2018115022.}
		\thanks{S. Buzzi, C. D'Andrea and A. Zappone are with the Department
			of Electrical and Information Engineering, University of Cassino and Southern Latium, Cassino,
			Italy, and with Consorzio Nazionale Interuniversitario per le Telecomunicazioni (CNIT), Parma, Italy.
			M. Fresia is with Huawei Technol. Duesseldorf GmbH Wireless Terminal Chipset Technology Lab, Munich, Germany. Y. Zhang  and Shulan Feng are with HiSilicon Technologies Balong Solution Department Bejing, China}}

	\maketitle
	
	\begin{abstract}
		Reconfigurable Intelligent Surfaces (RISs) are recently attracting a wide interest due to their capability of tuning wireless propagation environments in order to increase the system performance of wireless networks. 
		In this paper, a multiuser wireless network assisted by a RIS is studied and resource allocation algorithms are presented for several scenarios. 
		First of all, the problem of channel estimation is considered, and an algorithm  that permits separate estimation of the mobile user-to-RIS and RIS-to-base stations components is proposed. Then, 
		for the special case of a single-user system, three possible approaches are shown in order to optimize the Signal-to-Noise Ratio with respect to the beamformer used at the base station and to the RIS phase shifts.   {Then, for a multiuser system with two cells, assuming channel-matched beamforming, the geometric mean of the downlink Signal-to-Interference plus Noise Ratios across users is maximized with respect to the base stations transmit powers and RIS phase shifts configurations. 
			In this scenario, the RIS is placed at the cell-edge and some users are jointly served by two base stations to increase the system performance.}
		Numerical results show that the proposed procedures are effective and that the RIS brings substantial performance improvements to wireless system. 
	\end{abstract}
	
	\begin{IEEEkeywords}
		reconfigurable intelligent surfaces, resource allocation, MIMO, multicell systems 
	\end{IEEEkeywords}

	%
	\IEEEpeerreviewmaketitle

	\section{Introduction}
	While Massive multiple input multiple output (MIMO) has been a breakthrough technology that has significantly contributed to the evolution of wireless networks in the last decade and is currently being deployed worldwide by several telecom operators, new technologies and solutions have recently started to appear and gather attention as possible evolution of massive MIMO systems. These  include, among others, cell-free massive MIMO systems \cite{ngo2015cell,BuzziWCL2017}, the use of massive MIMO arrays for joint communication and sensing \cite{BuzziAsilomar2019}, large distributed antenna arrays \cite{amiri2018extremely}, and, also, reconfigurable intelligent surfaces (RISs) \cite{hu2017potential,hu2018beyond,di2019smart}. 
	RISs are thin surfaces that can be used to coat buildings, ceilings, or other surfaces; they have electromagnetic properties that can be tuned electronically through a software controller, and their use permits modifying the propagation environment of wireless signals, so as to be able to concentrate information signals where needed and thus to improve the 
	Signal-to-Interference plus Noise Ratio (SINR). RIS' elements are passive, in the sense that they have no RF chains and transmit and receive antennas; they just add a tunable phase offset to the impinging and reflected waves. 
	Prototypes of reconfigurable metasurfaces are currently being developed in many parts of the world
	\cite{VISOSURF_project,NTT_DoCoMo_LIS2019}.
	
	\noindent
	
	Recent surveys and tutorials on RIS-based communications have appeared in \cite{JSAC_RIS,Liaskos_COMMAG,RuiZhang_COMMAG,BasarMag2019,HuangMag2020}, where the fundamentals, main research results, and future lines of research of RIS-based systems are discussed.   {Innovative and emerging RIS applications 	include multicell networks\cite{Pan_MulticellNetworks_TWC2020}, simultaneous wireless information and power transfer\cite{Pan_SWIPT_JSAC2020},
		mobile edge computing networks\cite{Bai_MEC_JSAC2020}, multicast networks\cite{Zhou_Multicas_TSP2020}, physical layer security 	systems\cite{Hong_PLS_TCOM2020} and cognitive radio networks\cite{Zhang_Cognitive_TVT2020}.}
	
	In the following, we direct our attention on the issues of resource allocation and channel estimation in RIS-based wireless networks. 
	
	In \cite{EE_RISs}, the rate and energy efficiency of RIS-based multiple input single output (MISO) downlink systems are optimized with respect to the base station transmit powers and to the RIS phase shifts. The optimization is carried out by means of the alternating optimization, fractional programming, and sequential optimization frameworks. In \cite{Wu2018}, a similar scenario is addressed, with the difference that the problem of power minimization subject to minimum rate constraints is tackled. Also in this case, the tool of alternating optimization is used to allocate the base station transmit power and the RIS phase shifts. A RIS-based MISO downlink system is also analyzed in \cite{Yang2019}, assuming that the orthogonal frequency division multiplexing (OFDM) transmission scheme is considered. In \cite{Yu2019}, algorithms are devised for the maximization of the sum-rate in an RIS-based MISO system. Alternating optimization is employed to optimize the transmit beamformer and the RIS phase shifts. Alternating optimization methods are also used in \cite{Guo2019} to address the problem of sum-rate maximization in a RIS-based MISO downlink system. The phase shifts applied by the RIS are assumed to take on discrete values and the base station beamformer and the RIS phase shifts are optimized. In \cite{Jiang2019}, over-the-air computations in a multi-user RIS-based MISO channel is considered and  alternating optimization is merged with difference convex (DC)-programming to develop a method that is shown to outperform the traditional use of semi-definite relaxation methods. In \cite{Li2019b} a massive MIMO system aided by the presence of multiple RISs is considered. Assuming that a large number of reflectors is equipped on each RIS, the problem of maximizing the minimum signal-to-interference-plus-noise-ratio at the users is tackled with respect to the transmit precoding vector and the RISs phase shifts. In \cite{Liu2019}, a RIS with a discrete phase resolution is assumed, and the problem of precoding design in an RIS-based multi-user MISO wireless system is investigated. Rate maximization in a RIS-assisted MIMO link is tackled in \cite{Ning2019b}, assuming that the RIS is used to aid the communication between the transmitter and the receiver. In \cite{Han2019b}, power control for physical-layer broadcasting in RIS-empowered networks is discussed, with the constraints of quality of service for the mobile users. In \cite{Pan_MulticellNetworks_TWC2020}, a multi-cell scenario is considered, and a RIS is deployed at the boundary between multiple cells to aid cell-edge users. In this context, the  problem of weighted sum-rate maximization is tackled by alternating optimization of the base station transmit powers and of the RIS phase shifts. A single-user, RIS-based MISO system using millimeter waves transmissions is considered in \cite{Wang2019b}, studying the problem of transmit beamforming and RIS phase shifts allocation considering the presence of both a single RIS and multiple RISs. 
	
	In \cite{You2019} the problem of joint channel estimation and sum-rate maximization is tackled in the uplink of a single-user RIS-based system. First, a channel estimation method based on the discrete Fourier transform and on a truncation of Hadamard matrices is developed, and then rate optimization is tackled by  a low-complexity successive refinement algorithm. In \cite{Nadeem2019}, the minimum SINR achieved by linear detection in downlink RIS-based systems is characterized considering line-of-sight between the base station and the RIS, with the corresponding channel being either full-rank or unit-rank. In \cite{Nadeem2019b}, a minimum mean square error channel estimation method is devised for RIS-based networks and, based on the estimated channels, the RIS phase shifts are optimized by a gradient-based algorithm. In \cite{He2019}, the problem of channel estimation in RIS-based systems is addressed by developing a method for the estimation of the cascaded Tx-RIS and RIS-Rx channels.
	A novel RIS architecture is proposed in \cite{alexandropoulos2020hardware}, wherein, based on the existence of an active RF chain at the RIS, explicit channel estimation at the RIS side is realized. 
	Uplink channel estimation is RIS-based wireless networks is discussed in \cite{Chen2019b}, where the cascade channel including the channel from the transmitter to the RIS and from the RIS to the receiver is estimated by compressive sensing techniques.  In \cite{Lin2019} channel estimation for RIS-based networks is approached as a constrained estimation error minimization problem, that is tackled by the Lagrange multipliers and dual ascent-based  schemes. In \cite{Ning2019}, beam training is used for channel estimation in a massive MIMO, RIS-assisted wireless system operating in the THz regime.   {Reference \cite{Wang2019d} proposes a three-phase pilot-based channel estimation framework in which the channels involved in the communication between RIS, BS, and users are estimated switching on and off the RIS elements.} In \cite{cui2019efficient}, a low-complexity channel estimation method for RIS-based communication systems using millimeter waves is developed exploiting the sparsity of the millimeter wave channel and the large size of the RIS. In \cite{wei2020parallel}, channel estimation in the downlink of a RIS-assisted MISO communication system is devised using the alternating least squares algorithm to iteratively estimate the channel between the base station and the RIS, as well as the channels between the RIS and the users. In \cite{liu2019matrix}, channel estimation for a RIS-based multiuser MIMO system is formulated as a message-passing algorithm that enables to factorize the cascaded channels from transmitter to receiver. In \cite{Han2019}, a robust approach to the design of RIS-based networks is proposed, thus reducing the overhead required for channel estimation. However, robust and statistical approaches do not allow to exploit the reconfiguration capabilities of RISs, which enables to dynamically compensate for the random fluctuations of wireless channels.   {References \cite{Zhou_Robust_WCL2020,Zhou_Robust_TSP2020} consider robust beamforming based on the imperfect cascaded BS-RIS-user channels at the transmitter, formulating transmit power minimization problems subject to worst-case rate constraints corresponding to the bounded CSI error model.} In \cite{ZapTWC2020}, a model is developed to quantify the overhead required for channel estimation in RIS-based networks, and for deploying an optimized phase configuration on a RIS. Next, based on this model, and overhead-aware optimization of RIS-based systems is performed, for the maximization of the system rate, energy efficiency, and their trade-off. It is shown that, despite the overhead, resource allocation based on the instantaneous channel realizations, performs better than other allocations, provided the transmit and receive antennas do not grow too large. Similar results are obtained in \cite{ZapArXiv2020}, with reference to the problem of optimizing the number of reflectors to activate at the RIS, for the maximization of the rate, energy efficiency, and their trade-off.

	Following on this track, this paper considers resource allocation problems for several instances of wireless networks assisted by a RIS. The contribution of this paper can be summarized as follows.   {First of all, we tackle the problem of channel estimation and develop a protocol that permits estimating decoupled channel coefficients. The proposed protocol is based on the transmission of pilot signals from the mobile stations (MSs) and allows to compute the channel impulse response for any arbitrary RIS configuration.} This feature of the proposed channel estimation algorithm is crucial in order to enable practical implementation of the proposed resource allocation procedures.   {Then, we consider the special case of a single-cell network with a single-user and focus on maximizing the signal-to-noise ratio (SNR) with respect to the beamformer at the BS and the phase shifts at the RIS.} Three different algorithms are proposed to this end, one based on a classical alternating-maximization approach, and two based on the maximization of a lower bound and of an upper bound of the SNR. For the latter two cases, the optimal solution is obtained in closed form, and this enables a fast and computationally-efficient computation of the sought solution.  
	Finally, for a multi-user multi-cell system, we consider a scenario where some of the users may be jointly served by multiple BSs to improve performance, and maximize the geometric mean of the   {downlink SINRs} with respect to the transmit powers and to the RIS phases, 
	using the gradient algorithm and the alternating maximization methodology. 
	One further distinguishing feature of our study is that we consider the general cases that, for each MS, the direct BS-MS and the reflected BS-RIS-MS paths may be or may not be simultaneously active, whereas in many of the previously mentioned papers the assumption that the direct BS-MS path is blocked is necessary in order to solve the considered optimization problems.
	Numerical results will show that the proposed resource allocation procedures provide substantial performance improvements, as well as that they blend well with the proposed channel estimation procedures. 
	
	This paper is organized as follows. Next section is devoted to the description of the system model. Section III and IV contain the description of the proposed optimization procedures, for the single-user and the multiuser case, respectively.
	In Section V  numerical results are presented, showing the effectiveness of the proposed procedures, while, finally concluding remarks are given in Section VI.
	
	\subsection*{Notation}
	We use non-bold letters for scalars, $a$ and $A$, lowercase boldface letters, $\mathbf{a}$, for vectors and uppercase lowercase letters, $\mathbf{A}$, for matrices. The transpose, the inverse and the conjugate transpose of a matrix $\mathbf{A}$ are denoted by $\mathbf{A}^T$, $\mathbf{A}^{-1}$ and $\mathbf{A}^H$, respectively. The trace and the main diagonal of the matrix $\mathbf{A}$ are denoted as tr$\left(\mathbf{A}\right)$ and diag$\left(\mathbf{A}\right)$, respectively. The diagonal matrix obtained by the scalars $a_1,\ldots, a_N$ is denoted by diag$( a_1,\ldots, a_N)$. The $N$-dimensional identity matrix is denoted as $\mathbf{I}_N$, the $(N \times M)$-dimensional matrix with all zero entries is denoted as $\mathbf{0}_{N \times M}$ and $\mathbf{1}_{N \times M} $ denotes a $(N \times M)$-dimensional matrix with unit entries. The vectorization operator is denoted by vec$(\cdot)$ and the Kronecker product is denoted by $\otimes$. Given the matrices $\mathbf{A}$ and $\mathbf{B}$, with proper dimensions, the horizontal concatenation is denoted by $\left[\mathbf{A}, \mathbf{B}\right]$. The $(m,\ell$)-th entry  and the $\ell$-th column of the matrix $\mathbf{A}$ are denoted as $\left[\mathbf{A}\right]_{(m,\ell)}$ and $\left[\mathbf{A}\right]_{(:,\ell)}$, respectively. The block-diagonal matrix obtained from matrices $\mathbf{A}_1, \ldots, \mathbf{A}_N$ is denoted by blkdiag$\left( \mathbf{A}_1, \ldots, \mathbf{A}_N\right)$. The statistical expectation operator is denoted as $\mathbb{E}[\cdot]$; $\mathcal{CN}\left(\mu,\sigma^2\right)$ denotes a complex circularly symmetric Gaussian random variable with mean $\mu$ and variance $\sigma^2$.

	\begin{figure}[!t]
		\centering
		\includegraphics[scale=0.7]{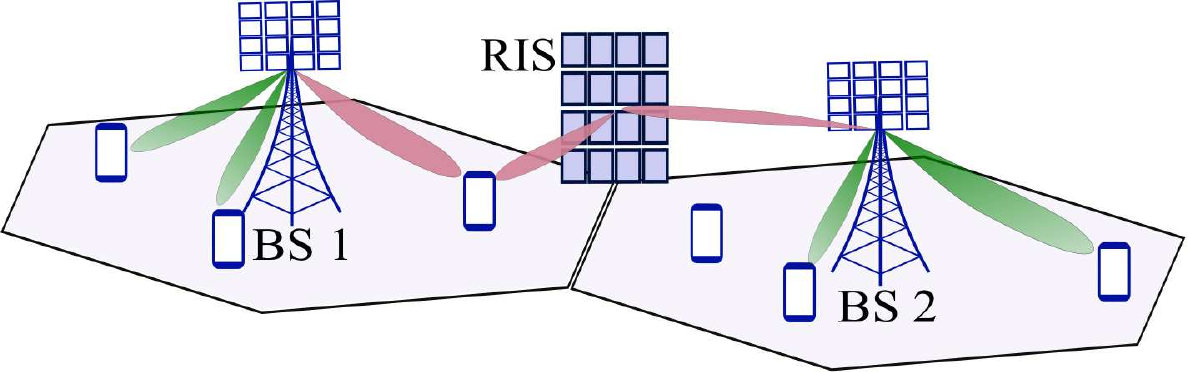}
		\caption{RIS-assisted multicell wireless network. Two BSs serve a set of users in the same frequency band relaying both on a direct link and on a further link reflected by a RIS, a planar array of reflecting devices with tunable phase reflection shift.}
		\label{Fig:Scenario_multi_BS}
	\end{figure}

	\section{System model}
	We consider a wireless cellular network and focus in particular on a system formed by two BSs,  equipped with $N_{B,1}$ and $N_{B,2}$ antennas, respectively. The two BSs may cooperate to serve $K$ single antenna mobile stations (MSs)  {\footnote{  {The cooperation between BSs is a well-investigated topic in literature and in reference \cite{Lozano_Cooperation_2013} the authors show that in presence of practical impairments, such as channel estimation and out of cell interference, it is essential to improve system performance.}}}, and both exploit a shared RIS with $N_R$ passive elements to improve the performance of the users\footnote{The extension to the case in which there are multiple RISs and more than two BSs can be   {done} with ordinary efforts and is not considered here for the sake of simplicity.} -- See Fig. \ref{Fig:Scenario_multi_BS}.

	We denote by $\widetilde{\bH}_1$ the $(N_{B,1} \times N_R)$-dimensional matrix representing the fast fading component of the wireless channel from the RIS to the first BS, by $\widetilde{\bh}_{k}$ the $N_R$-dimensional vector representing the fast fading component of the wireless channel from the $k$-th MS to the RIS, and by $\widetilde{\bh}_{1,k}^{(d)}$ the $N_{B,1}$-dimensional vector representing the fast fading component of the wireless direct link channel from the $k$-th MS to the first BS. Similarly $\widetilde{\bH}_2$ denotes the $(N_{B,2} \times N_R)$-dimensional matrix representing the fast fading component of the wireless channel from the RIS to the second BS and by $\widetilde{\bh}_{2,k}^{(d)}$ the $N_{B,2}$-dimensional vector representing the fast fading component of the wireless direct link channel from the $k$-th MS to the second BS. The RIS behaviour is modelled through a set of $N_R$ complex coefficients, representing the loss and phase shifts imposed on the reflected waves by each RIS elements. These coefficients are compactly represented through the diagonal matrix 
	$
	\bPhi= \textrm{diag}\left(\rho e^{j\phi_1}, \ldots, \rho e^{j\phi_{N_R}}\right)
	$
	The positive real-valued constant $\rho$ accounts for possible reflection losses and it is assumed to be constant across the RIS elements, while the phase offsets $\phi_1, \ldots, \phi_{N_R}$ can be controlled via software.
	
	Based on the above notation, the composite uplink channel from the $k$-th MS to the $i$-th BS (with $i \in \{1,2\}$) when both the direct link and the RIS-reflected link exist, can be easily shown to be written as
	\begin{equation}
		\overline{\bh}_{i,k, \bPhi}=\bH_i \bPhi \bh_k + \bh_{i,k}^{(d)} \, ,
		\label{comp_channel}
	\end{equation}
	where, in order to give evidence to the distance-dependent path-loss, we also let
	\begin{equation}
		\bH_i \bPhi \bh_k= \sqrt{\beta_{i,k}}\widetilde{\bH}_i \bPhi \widetilde{\bh}_k \; , \; \bh_{i,k}^{(d)}=\sqrt{\beta_{i,k}^{(d)}}\widetilde{\bh}_{i,k}^{(d)} \, ,
	\end{equation}
	where $\beta_{i,k}$ and $\beta_{i,k}^{(d)}$ denote the $k$-th MS power attenuation coefficients for the downlink RIS-reflected and direct links from the $i$-th BS, respectively.
	
	The Time Division Duplex (TDD) protocol is used to multiplex uplink (UL) and downlink (DL) data on the same carrier, thus implying that the UL and DL channels coincide.  The channel coherence time must be thus used to perform UL training for CE, DL data transmission, and UL data transmission. 
	
	\subsection{Signal model during UL training}
	Let us now focus on the signal model during the UL training phase.
	Denote by $\bp_k$ the unit energy vector pilot sequence assigned to the $k$-th MS, by $\tau_p < \tau_c$ the length (in discrete time samples) of the UL training phase and by $\tau_c$ the length
	(again in discrete time) of the coherence interval. During the UL training phase, each MS transmits its own pilot sequence; the discrete-time version of the baseband equivalent of the signal received at the $i$-th BS, with $i \in \{1,2\}$, can be thus represented through the following $(N_{B,i} \times \tau_p)$-dimensional matrix:
	\begin{equation}
		\bY_i=\ds \sum_{k=1}^K \sqrt{\eta_k} \overline{\bh}_{i,k, \bPhi} \bp_k^H + \bW_i \; .
		\label{eq:Y}
	\end{equation}
	In the above equation, the power coefficient $\eta_k$ is defined as $\eta_k=\tau_p \bar{\eta}_k$, with $\bar{\eta}_k$ the power transmitted by the $k$-th MS for each UL training symbol,  and the $(N_{i,B} \times \tau_p)$-dimensional matrix $\bW_i$ represents the additive white Gaussian noise (AWGN) contribution. It is assumed that the entries of $\bW_i$ are independent and identically distributed as ${\cal CN}(0, \sigma^2_w)$, with $\sigma^2_w$ the thermal noise variance. 
	
	\subsection{Signal model for DL data transmission}
	Focusing now on  the DL data transmission phase, let $\eta_{i,k}^{\rm DL}$ denote the DL transmit power
	reserved for the $k$-th MS by the $i$-th BS, with $i \in \{1,2\}$, and  let $q_k^{\rm DL}$ denote the information symbol to be delivered to the $k$-th MS. We consider the case in which the two BS cooperate and perform a joint transmission to the users that are located at the cell-edge. To this end, we define the following $\{0,1\}$-valued variate:
	\begin{equation}
		I_{i,\ell}=\left \lbrace
		\begin{array}{llll}
			1 \quad \text{if} \; \text{the } \ell\text{-th user is served by the } i\text{-th BS}
			\\
			0 \quad \text{otherwise}
		\end{array}\right. \, .
		\label{I_definition}
	\end{equation} 
	Denoting by  $\mathbf{w}_{i,k}$ the $N_{B,i}$-dimensional beamforming vector used at the $i$-th BS for the signal intended to the $k$-th MS, the signal transmitted by the $i$-th BS in the generic symbol interval can be thus expressed as
	\begin{equation}
		\mathbf{s}_{i}^{\rm DL}= \ds \sum_{\ell=1}^K  I_{i,\ell} {\sqrt{\eta_{i,\ell}^{\rm DL}} \mathbf{w}_{i,\ell}  q_{\ell}^{\rm DL}} \, ,
	\end{equation}
	while the signal received at the $k$-th MS is 
	\begin{equation}
		r_{k}^{\rm DL}= \overline{\bh}_{1, k, \bPhi}^H \mathbf{s}_{1}^{\rm DL} + \overline{\bh}_{2, k, \bPhi}^H \mathbf{s}_{2}^{\rm DL} \, .
	\end{equation}
	It thus follows that the soft estimate of the data symbol $q_k^{\rm DL}$ at the $k$-th MS can be written as 
	\begin{equation}
		\begin{array}{lll}
			&\!\!\!\!\!\widehat{q}_k^{\rm DL}=\underbrace{\left(I_{1,k}\sqrt{\eta_{1,k}^{\rm DL}}\overline{\bh}_{1, k, \bPhi}^H \mathbf{w}_{1,k} + I_{2,k}\sqrt{\eta_{2,k}^{\rm DL}}\overline{\bh}_{2, k, \bPhi}^H \mathbf{w}_{2,k}  \right)q_k^{\rm DL}}_{\mbox{useful contribution}}\\ &+ \ds \underbrace{\sum_{\substack{\ell=1 \\ \ell \neq k}}^K {\left(I_{1,\ell}\sqrt{\eta_{1,\ell}^{\rm DL}} \overline{\bh}_{1, k, \bPhi}^H \mathbf{w}_{1,\ell}  + I_{2,\ell}\sqrt{\eta_{2,\ell}^{\rm DL}} \overline{\bh}_{2, k, \bPhi}^H \mathbf{w}_{2,\ell}\right) q_{\ell}^{\rm DL}}}_{\mbox{interference}} \\ &+ z_k \, ,
		\end{array}
		\label{DL_signal_k}
	\end{equation}
	where $z_k \sim \mathcal{CN}(0,\sigma^2_z)$ is the AWGN contribution in the generic symbol interval.

	Based on the above expression, we can thus define the DL SINR as reported in Eq. \eqref{DL_SINR} at the top of next page.
	\begin{figure*}
		\begin{equation}
			\text{SINR}_{k, \bPhi}^{\rm DL}= \frac{\left|I_{1,k}\sqrt{\eta_{1,k}^{\rm DL}}\left(\bH_1 \bPhi \bh_k + \bh_{1,k}^{(d)}\right)^H \mathbf{w}_{1,k} + I_{2,k}\sqrt{\eta_{2,k}^{\rm DL}}\left( \bH_2 \bPhi \bh_k + \bh_{2,k}^{(d)} \right)^H \mathbf{w}_{2,k}\right|^2}{\ds \sum_{\substack{\ell=1 \\ \ell \neq k}}^K {\left|I_{1,\ell}\sqrt{\eta_{1,\ell}^{\rm DL}}\left(\bH_1 \bPhi \bh_k + \bh_{1,k}^{(d)}\right)^H \mathbf{w}_{1,\ell} + I_{2,\ell}\sqrt{\eta_{2,\ell}^{\rm DL}}\left(\bH_2 \bPhi \bh_k + \bh_{2,k}^{(d)}\right)^H \mathbf{w}_{2,\ell}\right|^2}+ \sigma^2_z} \, .
			\label{DL_SINR}
		\end{equation}
		%
		
		\hrulefill
		
	\end{figure*}

	\section{Channel estimation procedures} \label{Section_CE}
	Let us now tackle the problem of channel estimation. Based on the observation of $\bY_i$ in \eqref{eq:Y}, and relying on the knowledge of the pilot sequences $\bp_1, \ldots, \bp_K$, the $i$-th BS is faced with the task of performing the channel estimation (CE). Since this phase is implemented locally at the BSs without any cooperation, in the following we focus on the processing at the generic BS, and omit the subscript $i$ in order to simplify the notation.   {Since, in this paper, the RIS is a completely passive device controlled by the BSs, we only entrust the BSs with the task of estimating the channels involved in the communication.} It is worth noting that the unknown quantities to be estimated here are the $(N_B \times N_R)$-dimensional RIS-BS channel, the vectors containing the RIS-MS and BS-MS channels, for all the MSs in the system. A moment's thought however reveals that, given the model in Eq. \eqref{eq:Y} and the definition in Eq. \eqref{comp_channel}, the matrix $\bH$ and the channels  $\bh_{k}$ cannot be individually estimated; indeed, these quantities always appear as a product, thus implying that what can be actually estimated is the componentwise product between the rows of $\bH$ and the vectors $\bh_{k}$. 
	Luckily, this is enough to be able to predict the overall channel response when the RIS configuration changes. 
	More precisely, given the identity 
	\beq
	\bH \bPhi \bh_k= \bD_k  \bm{\phi}\; ,
	\label{eq:identity}
	\eeq
	where $\bD_k(i,j) \triangleq \bH(i,j) \bh_k(j)$, for all $i=1, \ldots, N_B$, $j=1, \ldots, N_R$ and $\bm{\phi}=\textrm{diag}(\bPhi)$ is the  $N_R$-dimensional column vector containing the diagonal entries of $\bPhi$, the matrix $\bY$ in Eq. \eqref{eq:Y} is easily shown to contain a linear combination of the channel coefficients.
	For the estimation of the composite channel for the generic $k$-th MS, we focus on the $N_B$-dimensional observable $\overline{\by}_k=\bY \bp_k/\sqrt{\eta_k}$. Using Eqs. \eqref{comp_channel}, \eqref{eq:Y} and \eqref{eq:identity}, $\overline{\by}_k$ can be written as
	\begin{equation}
		\begin{array}{llll}
			\overline{\by}_k= &\ds \bD_k \bm{\phi} + \bh_{k}^{(d)} + \overline{\bw}_k \\ &+ \ds \sum_{\substack{j=1 \\ j \neq k}}^K \sqrt{\frac{\eta_j}{\eta_k}}\left( \bD_j \bm{\phi}  +
			\bh_{j}^{(d)}\right) p_{j,k}\; ,
		\end{array}
		\label{y_k1}
	\end{equation}
	with $\overline{\bw}_k=\bW \bp_k \sim \mathcal{CN}\left(0, \sigma^2_w \mathbf{I}_{N_B}\right)$ and $p_{j,k}=\bp_j^H \bp_k$.
	Given Eq. \eqref{y_k1}, the CE problem is now well defined as the problem of estimating the $(N_B \times N_R)$-dimensional matrices $\bD_k$ and the $N_B$-dimensional vectors $ \bh_{k}^{(d)}$ for all $k=1, \ldots K$. 
	In order to conveniently cast the problem of CE, we write the noiseless part of the observables as the product of a known matrix times a column vector containing all the unknowns to be estimated. Otherwise stated, letting 
	$
	\widetilde{\by}= \left[ \begin{array}{llll}
		\overline{\by}_1^T, \; \ldots , \; \overline{\by}_K^T
	\end{array} \right]^T
	$
	be the $K N_B$-dimensional observable containing the projection of the received data $\bY$ onto the 
	pilot sequences $\bp_1, \ldots, \bp_K$, 
	upon defining the vector $N_B(N_R+1)$-dimensional vector $\widetilde{\mathbf{d}}_k$ as
	$$
	\widetilde{\mathbf{d}}_k= \left[
	\text{vec}( \bD_k); 
	\bh_{k}^{(d)}\right] \, ,
	$$
	the $K N_B(N_R+1)$-dimensional vector $
	\mathbf{d}= \left[ \widetilde{\mathbf{d}}_1^T, \; \ldots ,\; \widetilde{\mathbf{d}}_K^T \right]^T 
	$, and letting $\widetilde{\bA}$ be a matrix whose  $(k,j)$-th block, say $\widetilde{\bA}_{k,j}$, 
	with dimension $[N_B \times N_B(N_R+1)]$,
	is defined as 
	\begin{equation}
		\widetilde{\bA}_{k,j}= \sqrt{\frac{\eta_j}{\eta_k}} p_{j,k}\left[ \bm{\phi}^T \otimes \mathbf{I}_{N_B}, \mathbf{I}_{N_B} \right]\; ,
	\end{equation}
	for all $k, j = 1, \ldots, K$, then it can be shown that 
	the observable $\widetilde{\by}$ can be expressed as follows
	\begin{equation}
		\widetilde{\by}= \widetilde{\bA} \mathbf{d} + \widetilde{\mathbf{w}} \; .
		\label{y_tilde}
	\end{equation}
	Expression \eqref{y_tilde} reveals the linear relationship between the data $\widetilde{\by}$ and the $K N_B(N_R+1)$-dimensional vector 
	$\mathbf{d}$ to be estimated. In the following, we outline two possible CE strategies.

	\subsection{Least squares (LS) CE} \label{LS_CE_Section}
	Since in \eqref{y_tilde} the number of parameters to be estimated is larger than the number of available measurements, a least squares procedure cannot be directly applied. To circumvent this problem,  we assume that the pilot sequences are transmitted by the MSs $Q$ times, each one with a different RIS configuration. This assumption is needed since we should generate a number of observables that is not smaller than the number of unknown coefficients in order to be able to use a linear estimation rule. 
	We thus assume that the RIS phase shifts assume $Q$ different configurations, and denote by $$\bPhi^{(q)}= \textrm{diag}\left(\rho e^{j\phi_1^{(q)}}, \ldots, \rho e^{j\phi_{N_R}^{(q)}}\right) \; , \forall q=1,\ldots,Q, $$ 
	the diagonal matrix representing the RIS in the $q$-th configuration. Letting $\widetilde{\by}^{(q)}$ denote the 
	$(K N_B)$-dimensional observable vector when the RIS is in the $q$-th state, we form the following $QN_B$-dimensional observable
	\begin{equation}
		\widetilde{\by}_Q= \left[ \begin{array}{llll}
			\widetilde{\by}^{(1)\, T},  \; \ldots,  \; \widetilde{\by}^{(Q)\, T}
		\end{array} \right]^T= \widetilde{\bA}_Q \mathbf{d} + \widetilde{\mathbf{w}}_Q \; ,
		\label{y_tilde_Q}
	\end{equation}
	with $\widetilde{\mathbf{w}}_Q= \left[\widetilde{\mathbf{w}}^{(1) \, T}, \, \ldots ,\, \widetilde{\mathbf{w}}^{(Q), \, T} \right]^T$ and
	$
	\widetilde{\bA}_Q= \left[  \widetilde{\bA}^{(1) \, T}\, , \ldots , \; \widetilde{\bA}^{(Q)\, T} \right]^T
	$.
	The LS-based estimator of $\mathbf{d}$ can be thus simply written as
	\begin{equation}
		\widehat{\mathbf{d}}_{\rm LS} =\widetilde{\bA}_Q^{-1}\widetilde{\by}_Q \, .
		\label{d_estimation_LS}
	\end{equation}
	The number of needed RIS configurations in order to be able to correctly implement Eq. \eqref{d_estimation_LS} is $Q=N_R+1$.

	\subsection{Linear minimum mean square error (MMSE) CE} \label{MMSE_CE_Section}
	Another possible approach is based on the use of linear MMSE estimation. To this end, 
	we assume perfect knowledge of the large-scale power attenuation coefficients for all the MSs, i.e. the quantities  $\beta_k$ and $\beta_{k,d}, \; \forall \; k=1,\ldots,K$ are known at the BS.
	For linear MMSE estimation, the channel can be estimated even with just one configuration of the RIS elements. In the following, we thus consider both the cases that the number $Q$ of RIS configurations is larger than 1 and that is exactly 1. 
	
	\subsubsection{Linear MMSE CE with  $Q>1$  configurations of the RIS} 
	Based on \eqref{y_tilde_Q}, the linear MMSE estimate for the vector $\mathbf{d}$ can be written as \cite{kay1993fundamentals}
	\begin{equation}
		\widehat{\mathbf{d}}_{\rm MMSE, Q} =\mathbf{E}_Q^H\widetilde{\by}_Q \, ,
		\label{d_estimation_MMSE}
	\end{equation}
	with $\mathbf{E}_Q$ is a suitable $\left[K N_B Q \times K N_B (N_R+1)\right]$-dimensional matrix, such that  the statistical expectation of the squared estimation error  $\|\widehat{\mathbf{d}}_{\rm MMSE, Q} - \mathbf{d}\|^2$ is minimized. 
	Applying well-known statistical signal processing results \cite{kay1993fundamentals}, we have:
	\begin{equation}
		\mathbf{E}_Q=\left( \widetilde{\bA}_Q \mathbf{R}_{d} \widetilde{\bA}_Q^H + \sigma^2_w \mathbf{I}_{K N_B Q}\right)^{-1} \widetilde{\bA}_Q \mathbf{R}_{d} \, ,
		\label{E_Q}
	\end{equation}
	where $\mathbf{R}_{d}=\mathbb{E}\left[ \mathbf{d} \mathbf{d}^H \right]=\text{blkdiag}\left( \mathbf{R}_{d}^{(1)}, \ldots,  \mathbf{R}_{d}^{(K)}\right)$, 
	with
	\begin{equation}
		\mathbf{R}_{d}^{(k)}=\text{diag}\left( \underbrace{\beta_k, \ldots, \beta_k}_{N_B N_R}, \underbrace{\beta_{k,d},\ldots, \beta_{k,d}}_{N_B}\right) \; ,
		\label{R_d_twolinks}
	\end{equation}
	
	\subsubsection{Linear MMSE CE with $Q=1$ RIS configuration} 
	With regard to the case in which the RIS assumes just one configuration and the pilot sequences are trasnmitted by each MS just one, the linear MMSE estimator for the channel coefficients can be obtained by simply specializing to the case $Q=1$ the derivations of Eq. \eqref{d_estimation_MMSE}. We omit providing further details for the sake of brevity.

	\section{Joint beamformer and RIS configuration design in a single-user system} \label{Single_user_Resource}
	This section focuses on the special case of a single-user system served by just one BS, which may be representative of a network with an orthogonal multiple access scheme and with negligible co-channel interference. 
	We tackle the maximization of the system SNR with respect to the base station beamforming vector $\mathbf{w}$ (active beamforming) and of the RIS phase shifts (which we refer to as passive beamforming). In agreement with the previously defined notation, in the following we denote by $\eta^{\rm DL}$ the BS transmit power during the data transmission phase, by $q^{\rm DL}$ the information symbol intended for the MS in the generic (discrete) symbol interval. The soft estimate of such data symbol at the MS  can be expressed as 
	\begin{equation}
		\begin{array}{lll}
			\widehat{q}^{\rm DL}=\sqrt{\eta^{\rm DL}} \left( \bH \bPhi \bh +\bh^{(d)} \right)^H \mathbf{w}  q^{\rm DL} + z \, ,
		\end{array}
		\label{DL_signal_k_SU}
	\end{equation}
	with $z \sim \mathcal{CN}(0,\sigma^2_z)$ denoting thermal noise.
	
	Our first step, is to rewrite \eqref{DL_signal_k_SU} in a more convenient form, by exploiting the  identity in Eq. \eqref{eq:identity}, which enables us to rewrite \eqref{DL_signal_k_SU} as
	\begin{equation}
		\begin{array}{lll}
			\widehat{q}^{\rm DL}=\sqrt{\eta^{\rm DL}} \left( \bD  \bm{\phi} +\bh^{(d)} \right)^H \mathbf{w}  q^{\rm DL} + z \, ,
		\end{array}
		\label{DL_signal_k2_SU}
	\end{equation}
	Based on \eqref{DL_signal_k_SU}, the system SNR can be defined as
	\begin{equation}
		\text{SNR}= \frac{\eta^{\rm DL}}{\sigma_z^2} \left| \mathbf{w}^H \left( \bD  \bm{\phi} +\bh^{(d)} \right) \right|^2
	\end{equation}
	In practice, the BS will have access to estimates of $\bD$ and $\bh^{(d)}$, which will be denoted in the following by $\widehat{\bD}$ and $\widehat{\bh}^{(d)}$, respectively, which implies that the function that can be optimized at the transmit side is 
	\begin{equation}
		\widehat{\text{SNR}}= \frac{\eta^{\rm DL}}{\sigma_z^2} \left| \mathbf{w}^H \left( \widehat{\bD}  \bm{\phi} +\widehat{\bh}^{(d)} \right) \right|^2
	\end{equation}
	Then, the problem to solve is stated as 
	\begin{subequations}\label{Prob:MaxSNR}
		\begin{align}
			&\ds\max_{\mathbf{w}, \bm{\phi}}\; \; \; \;  \left| \mathbf{w}^H \left( \widehat{\bD}  \bm{\phi} +\widehat{\bh}^{(d)} \right) \right|^2  \label{Prob:Max_SNR}\\
			&\;\textrm{s.t.}\; \; [\bm{\phi}]_i= \rho e^{j \phi_{i}},  \\
			&\;\; \; \;\; \; \; \phi_{i} \, \in \, [-\pi, \pi], \, \forall \; i=1, \ldots, N_R \\
			&\;\; \;\;\; \; \; \|\mathbf{w}\|^2=1
		\end{align}
	\end{subequations}
	{Problems} of the form of \eqref{Prob:MaxSNR} are usually tackled by alternating optimization methods which iterate between the optimization of the base station beamforming vector $\mathbf{w}$ and of the RIS phase shifts $\bm{\phi}$. This approach could be used also for the case at hand, but it has the drawback of requiring a numerical iterative algorithm. Instead, in the following, we propose two optimization methods that optimize an upper-bound and a lower-bound of the objective of \eqref{Prob:MaxSNR}, and which have the advantage of leading to closed-form expressions of $\mathbf{w}$ and $\bm{\phi}$. 
	
	\subsection{Upper-bound maximization} \label{UB_max_Section}
	Assume, without loss of generality, that $N_B<N_R$, and consider the singular value decomposition of $\widehat{\bD}$, i.e.,
	\begin{equation}
		\widehat{\bD}=\sum_{i=1}^{N_B} \lambda_i \mathbf{u}_i \mathbf{v}_i^H\;.
	\end{equation}
	Next, let us express $\widehat{\bh}^{(d)}$ in terms of its projection on the orthonormal basis vectors $\mathbf{u}_1, \ldots, \mathbf{u}_{N_B}$, i.e., 
	\begin{equation}
		\widehat{\bh}^{(d)}=\sum_{i=1}^{N_B} \alpha_i \mathbf{u}_i, 
	\end{equation}
	where $\alpha_i=\mathbf{u}_i^H\widehat{\bh}^{(d)}$. 
	
	At this point, we observe that an upper bound of the objective of Problem \eqref{Prob:MaxSNR} can be written as
	\begin{equation}
		\begin{array}{lllll}
			\left| \mathbf{w}^H \left( \widehat{\bD}  \bm{\phi} +\widehat{\bh}^{(d)} \right) \right|^2 &= \left| \mathbf{w}^H \left[ \ds \sum_{i=1}^{N_B} \mathbf{u}_i \left( \lambda_i \mathbf{v}_i^H \bm{\phi} +\alpha_i\right) \right] \right|^2 \\ & \leq N_B \ds  \sum_{i=1}^{N_B} \left|\mathbf{w}^H\mathbf{u}_i\right|^2 \left| \lambda_i \mathbf{v}_i^H \bm{\phi} +\alpha_i \right|^2
		\end{array}
		\label{SNR}
	\end{equation} 
	The upper-bound in \eqref{SNR} can be jointly maximized with respect to both $\bm{\phi}$ and $\mathbf{w}$. To see this, let us first consider, for all $i=1,\ldots,N_{B}$, the following optimization problem
	\begin{equation}
		\ds\max_{\bm{\phi}} \left| \lambda_i \mathbf{v}_i^H \bm{\phi} +\alpha_i \right|^2 = \ds\max_{\bm{\phi} } \left| \lambda_i \mathbf{v}_i^H \bm{\phi} e^{-j \angle{\alpha_i}} +|\alpha_i| \right|^2\;,
		\label{max_phi_i_UB}
	\end{equation}
	whose optimal solution $\bm{\phi}_i^{\rm opt}$ is found by noticing that the phase of the $n$-th entry of $\bm{\phi}_i^{\rm opt}$, say $\phi_{n,i}^{\rm opt}$, is given by 
	\begin{equation}
		\phi_{n,i}^{\rm opt}= -\angle{[\mathbf{v}_i^*]_n} + \angle{\alpha_i}\;.
	\end{equation}
	Next, let us define the index $i^{+}=\text{argmax}_{i}\left| \lambda_i \mathbf{v}_i^H \bm{\phi}_i^{\rm opt} +\alpha_i \right|^2$ and $c_{i^{+}}=\left| \lambda_{i^{+}} \mathbf{v}_{i^{+}}^H \bm{\phi}_{i^{+}}^{\rm opt} +\alpha_{i^{+}}\right|^2$. Thus, it follows that 
	\begin{equation}
		\ds  \sum_{i=1}^{N_B} \left|\mathbf{w}^H\mathbf{u}_i\right|^2 \left| \lambda_i \mathbf{v}_i^H \bm{\phi} +\alpha_i \right|^2\leq 
		c_{i^{+}}\sum_{i=1}^{N_{B}}|\mathbf{w}^{H}\mathbf{u}_{i}|^{2}\leq c_{i^{+}}\;,\label{Eq:UpperBound}
	\end{equation}
	where we have also exploited that fact that both $\mathbf{w}^{H}$ and $\mathbf{u}_{i}$ have unit-norm. Finally, we observe that all inequalities in \eqref{Eq:UpperBound} turn to equalities by choosing 
	$
	\bm{\phi}^{\rm opt}=\bm{\phi}_{i^+}^{\rm opt}$ and $\mathbf{w}^{\rm opt}=\mathbf{u}_{i^+}$,
	which therefore are the maximizers of the right-hand-side of \eqref{SNR}. 
	
	\subsection{Lower-bound Maximization} \label{LB_Max_Section}
	Define $\mathbf{g}_{w}^H=\mathbf{w}^H \widehat{\bD}$, and $t_{w}=\mathbf{w}^H\widehat{\bh}^{(d)}$. Then, it holds:
	\begin{align}
		&\ds\max_{\mathbf{w},\bm{\phi}}\left|\mathbf{w}^{H}(\widehat{\bD}\bm{\phi}+\widehat{\bh}_{d})\right|^{2}= \ds \max_{\mathbf{w}}\left(\max_{\bm{\phi}}\left|\mathbf{g}_{w}^{H}\bm{\phi}+t_{w})\right|^{2}\right)\stackrel{(a)}{=}\notag\\
		&\ds\rho^2 \max_{\mathbf{w}}\left|\sum_{i=1}^{N_{R}}|\mathbf{g}_{w}(i)|+|t_{w}|\right|^{2}\stackrel{(b)}\geq
		\ds \rho^2\max_{\mathbf{w}}\left|\mathbf{w}^{H}\left(\sum_{i=1}^{N_{R}}\widehat{\mathbf{d}}_{i}+\widehat{\bh}^{(d)}\right)\right|^{2}
		\label{SNR_LB}
	\end{align} 
	where the equality $(a)$ holds since, for any given $\mathbf{w}$, the optimal $\bm{\phi}$ is the one that aligns the phases of $t_{w}$ and of the components of $\mathbf{g}_{w}$, denoted by $\mathbf{g}_{w}(i)$ with $i=1,\ldots,N_{R}$, while inequality (b) holds by removing the inner absolute values and since $\mathbf{g}_{w}(i)=\mathbf{w}^{H}\widehat{\mathbf{d}}_i$, with $\widehat{\mathbf{d}}_i$ the  $i$-th column of $\widehat{\bD}$.
	
	From \eqref{SNR_LB}, we see that the optimal $\mathbf{w}$ has the form:
	\begin{equation}
		\mathbf{w}^{\rm opt}=\frac{\ds \sum_{i=1}^{N_B} \widehat{\mathbf{d}}_i + \widehat{\bh}^{(d)}}{\norm{\ds \sum_{i=1}^{N_B} \widehat{\mathbf{d}}_i + \widehat{\bh}^{(d)}}}\;,
	\end{equation}
	from which we can obtain the optimal phases of the RIS as  
	\begin{equation}
		\phi_{i}^{\rm opt}= -\angle{\mathbf{g}_{w}^{*}(i)} + \angle{t_w}\;,\;\forall\;n=1,\ldots,N_{R} .
	\end{equation}
	
	\section{Joint RIS configuration design and power allocation in multiuser systems with joint transmission} \label{Joint_Resource}
	{In the general multi-user scenario, we assume that the users may be jointly served by the two BSs to improve performance, and tackle the problem of the maximization of the geometric mean of the downlink SINRs expressed as in Eq. \eqref{DL_SINR} with respect to the transmit powers and to the RIS phases. Exploiting the definition in Eq. \eqref{eq:identity} and rewriting Eq. \eqref{DL_SINR} in a more compact form the problem to solve is stated as follows:}
	
	\begin{subequations}\label{Prob:MaxSNR_MU}
		\begin{align}
			\ds\max_{\bm{\eta}_1^{\rm DL}, \bm{\eta}_2^{\rm DL}, \bm{\phi}}  &  \prod_{k=1}^K \! \frac{\ds \sum_{i=1}^2\left|I_{i,k}\sqrt{\eta_{i,k}^{\rm DL}}\left(\widehat{\bD}_{i,k}  \bm{\phi} + \widehat{\bh}_{i,k}^{(d)}\right)^H\!\!\! \mathbf{w}_{i,k}\right|^2}{\ds \sum_{\substack{\ell=1 \\ \ell \neq k}}^K {\left|
					\ds \sum_{i=1}^2
					I_{i,\ell}\sqrt{\eta_{i,\ell}^{\rm DL}}\left(\widehat{\bD}_{i,k}  \bm{\phi} + \widehat{\bh}_{i,k}^{(d)}\right)^H \!\!\!\!\mathbf{w}_{i,\ell}\right|^2}\!\!+\! \sigma^2_z}   , \\
			\textrm{s.t.} \quad &  [\bm{\phi}]_n= \rho e^{j \phi_{n}},  \\
			&   \phi_{n} \, \in \, [-\pi, \pi], \, \forall \; n=1, \ldots, N_R \\
			&   \sum_{\ell=1}^K { I_{i,\ell}\eta_{i,\ell}^{\rm DL}} \leq P_{\rm max}^{{\rm BS},i}, \; , i=1,2\\
			&  {\eta_{i,\ell}^{\rm DL}} \geq 0 \; \forall \ell=1,\ldots,K \; , i=1,2  
		\end{align}
	\end{subequations}
	where $\bm{\eta}_1^{\rm DL}= \left[ \eta_{1,1}^{\rm DL}, \ldots, \eta_{1,K}^{\rm DL}\right]^T$ , $\bm{\eta}_2^{\rm DL}= \left[ \eta_{2,1}^{\rm DL}, \ldots, \eta_{2,K}^{\rm DL}\right]^T$ and we are assuming again that the BSs treat the channel estimates as the true channels. Substituting $\bm{\phi}= \rho e^{j \bm{\widetilde{\phi}}}$, with $\bm{\widetilde{\phi}}= \left[ \phi_1, \ldots \ \phi_{N_R}\right]^T$ and assuming channel-matched beamforming (CM-BF), i.e., the beamforming vector at the $i$-th BS to the $k$-th MS is chosen as
	\begin{equation}
		\mathbf{w}_{i,k}= \displaystyle \frac{\rho \widehat{\bD}_{i,k} e^{j\bm{\widetilde{\phi}}}+\widehat{\bh}_{i,k}^{(d)}}{\norm{\rho \widehat{\bD}_{i,k} e^{j\bm{\widetilde{\phi}}}  +\widehat{\bh}_{i,k}^{(d)}}}  \; .
		\label{CM_BF}
	\end{equation} 
	Problem \eqref{Prob:MaxSNR_MU} can be rewritten as Problem \eqref{Prob:MaxSNR_MU2} at the top of next page.
	\begin{figure*} 
		\begin{subequations}\label{Prob:MaxSNR_MU2}
			\begin{align}
				\ds\max_{\bm{\eta}_1^{\rm DL}, \bm{\eta}_2^{\rm DL}, \bm{\widetilde{\phi}}} \; &  \prod_{k=1}^K \frac{
					\ds \sum_{i=1}^2\left|I_{i,k}\sqrt{\eta_{i,k}^{\rm DL}}\norm{\rho \widehat{\bD}_{i,k} e^{j \bm{\widetilde{\phi}}}  +\widehat{\bh}_{i,k}^{(d)}}\right|^2}{\ds \sum_{\substack{\ell=1 \\ \ell \neq k}}^K \left|
					\ds \sum_{i=1}^2 I_{i,\ell}\sqrt{\eta_{i,\ell}^{\rm DL}}\frac{\left(\rho \widehat{\bD}_{i,k} e^{j\bm{\widetilde{\phi}}}  +\widehat{\bh}_{i,k}^{(d)}\right)^{\!\!H}\!\!\!\!\left( \rho \widehat{\bD}_{i,\ell} e^{j\bm{\widetilde{\phi}}} \!\! +\!\!\widehat{\bh}_{i,\ell}^{(d)} \right)}{\norm{\rho \widehat{\bD}_{i,\ell} e^{j\bm{\widetilde{\phi}}}  +\widehat{\bh}_{i,\ell}^{(d)}}} \right|^2  \!\! + \!\!  \sigma^2_z} \, , \\
				\textrm{s.t.} \quad &   \left[\bm{\widetilde{\phi}}\right]_{n} \, \in \, [-\pi, \pi], \, \forall \; n=1, \ldots, N_R ,  \\
				&   \sum_{\ell=1}^K { I_{i,\ell}\eta_{i,\ell}^{\rm DL}} \leq P_{\rm max}^{{\rm BS},i}, \; , i=1,2\\
				&  {\eta_{i,\ell}^{\rm DL}} \geq 0 \; \forall \ell=1,\ldots,K \; , i=1,2.    
			\end{align}
		\end{subequations}
	\end{figure*}
	Solving \eqref{Prob:MaxSNR_MU2} optimally appears challenging, due to the presence of multi-user interference. This motivates us to resort to alternating optimization to find a candidate solution of  \eqref{Prob:MaxSNR_MU2} \cite[Section 2.7]{BertsekasNonLinear}, i.e, we solve alternatively the problem with respect to $\bm{\widetilde{\phi}}$ and then the problem with respect to $\bm{\eta}_1^{\rm DL}$ and $\bm{\eta}_2^{\rm DL}$. At each step, the objective of \eqref{Prob:MaxSNR_MU2} does not decrease, and so the algorithm converges in the value of the objective function.
	
	\subsection{Solution of the problem with respect to $\bm{\widetilde{\phi}}$} \label{Phi_opt_Section}
	Determining the optimal solution of Problem \eqref{Prob:MaxSNR_MU2} appears challenging even with respect to only the RIS phases, due to the fact multiple users are present, that are served by the same RIS matrix $\bm{\phi}$. This prevents from obtaining significant insight on the optimal $\bm{\Phi}$. On the other hand, it was shown in \cite{EE_RISs} that good results are obtained when employing a gradient-based search to optimize the phase shift matrix in RIS-based networks. Here we take a similar approach, applying the gradient algorithm in order to find a candidate solution for Problem \eqref{Prob:MaxSNR_MU2}. Before applying the gradient algorithm, we equivalently reformulate the problem by taking the logarithm of the objective, which leads us to Problem \eqref{Prob:MaxSNR_MU1_phi}, shown in the next page\footnote{Without loss of generality, we have neglected the constraint $\left[\bm{\widetilde{\phi}}\right]_{n} \, \in \, [-\pi, \pi], \, \forall \; n=1, \ldots, N_R$, since the objective is periodic with respect to each phase, with period $2\pi$, and thus any phase can be restricted to this fundamental period after the optimization routine has converged.}. 
	
	\begin{figure*}
		\begin{subequations}\label{Prob:MaxSNR_MU1_phi}
			\begin{align}
				&\ds\max_{\bm{\widetilde{\phi}}} \sum_{k=1}^K \log_2 \left(\frac{\left|
					\ds \sum_{i=1}^2
					I_{i,k}\sqrt{\eta_{i,k}^{\rm DL}}\norm{\rho \widehat{\bD}_{i,k} e^{j \bm{\widetilde{\phi}}}  +\widehat{\bh}_{i,k}^{(d)}} \right|^2}{\ds \sum_{\substack{\ell=1 \\ \ell \neq k}}^K \left|
					\ds \sum_{i=1}^2
					I_{i,\ell}\sqrt{\eta_{i,\ell}^{\rm DL}}\frac{\left(\rho \widehat{\bD}_{i,k} e^{j\bm{\widetilde{\phi}}}  +\widehat{\bh}_{i,k}^{(d)}\right)^{\!\!H}\!\!\!\!\left( \rho \widehat{\bD}_{i,\ell} e^{j\bm{\widetilde{\phi}}} \!\! +\!\!\widehat{\bh}_{i,\ell}^{(d)} \right)}{\norm{\rho \widehat{\bD}_{i,\ell} e^{j\bm{\widetilde{\phi}}}  +\widehat{\bh}_{i,\ell}^{(d)}}} \right|^2  \!\! + \!\!  \sigma^2_z} \right) \, , 
			\end{align}
		\end{subequations}
	\end{figure*}
	
	Solving \eqref{Prob:MaxSNR_MU1_phi} still appears challenging even with respect to only the RIS phases, due to the fact multiple users and base stations are present. This motivates the use of the gradient algorithm to find a candidate solution for $\bPhi$, which was shown to yield good results in \cite{EE_RISs}, although in a less challenging scenario as the one considered here. 
	
	To elaborate, let us define, for $i\in\{1,2\}$, 
	
	\begin{equation}
		F_{k,\ell}^{(i)}=I_{i,\ell}\eta_{i,\ell}^{DL}\left( \rho \widehat{\bD}_{i,k} e^{j\bm{\widetilde{\phi}}} +\widehat{\bh}_{i,k}^{(d)}\right)^{H}\left( \rho \widehat{\bD}_{i,\ell} e^{j\bm{\widetilde{\phi}}}  +\widehat{\bh}_{i,\ell}^{(d)} \right)\;.
		\label{F_k_ell_def}
	\end{equation}
	Then, denoting by $G(\bm{\widetilde{\phi}})$ the objective of \eqref{Prob:MaxSNR_MU1_phi}, it holds that 
	\begin{equation}
		G(\bm{\widetilde{\phi}})\!=\!\!\sum_{k=1}^{K}\!\log_{2}\!\!\left(\!\!\frac{F_{k,k}^{(1)}+F_{k,k}^{(2)}+2\sqrt{F_{k,k}^{(1)}F_{k,k}^{(2)}}}{\ds\sigma_{z}^{2}\!+\!\sum_{\ell\neq k}\frac{|F_{k,\ell}^{(1)}|^{2}}{F_{\ell,\ell}^{(1)}}\!+\!\frac{|F_{k,\ell}^{(2)}|^{2}}{F_{\ell,\ell}^{(2)}}\!+\!\frac{2\Re\left\{F_{k,\ell}^{(1)}F_{k,\ell}^{*(2)}\right\}}{\sqrt{F_{\ell,\ell}^{(1)}F_{\ell,\ell}^{(2)}}}}\!\!\right)
	\end{equation}
	Therefore, denoting by $\widetilde{\phi}_{n}$ the $n$-th component of $\bm{\widetilde{\phi}}$, for any $n=1,\ldots,N_{R}$, the derivative of $G$ with respect to $\widetilde{\phi}_{n}$ can be expressed as shown in Eq. \eqref{Eq:DerG} at the bottom of next page,  
	\begin{figure*}
		\begin{align}\label{Eq:DerG}
			&\frac{\partial G}{\partial \widetilde{\phi}_{n}}=\log_{2}(e)\sum_{k=1}^{K}\frac{\ds\sigma_{z}^{2}\!+\!\sum_{\ell\neq k}\frac{|F_{k,\ell}^{(1)}|^{2}}{F_{\ell,\ell}^{(1)}}\!+\!\frac{|F_{k,\ell}^{(2)}|^{2}}{F_{\ell,\ell}^{(2)}}\!+\!\frac{2\Re\left\{F_{k,\ell}^{(1)}F_{k,\ell}^{*(2)}\right\}}{\sqrt{F_{\ell,\ell}^{(1)}F_{\ell,\ell}^{(2)}}}}{F_{k,k}^{(1)}+F_{k,k}^{(2)}+2\sqrt{F_{k,k}^{(1)}F_{k,k}^{(2)}}}\\
			&\times\Bigg\{\left(\frac{\partial F_{k,k}^{(1)}}{\partial \widetilde{\phi}_{n}}+\frac{\partial F_{k,k}^{(2)}}{\partial \widetilde{\phi}_{n}}+2\frac{\partial \sqrt{F_{k,k}^{(1)}F_{k,k}^{(2)}}}{\partial \widetilde{\phi}_{n}}\right)\left(\sigma_{z}^{2}\!+\!\sum_{\ell\neq k}\frac{|F_{k,\ell}^{(1)}|^{2}}{F_{\ell,\ell}^{(1)}}\!+\!\frac{|F_{k,\ell}^{(2)}|^{2}}{F_{\ell,\ell}^{(2)}}\!+\!\frac{2\Re\left\{F_{k,\ell}^{(1)}F_{k,\ell}^{*(2)}\right\}}{\sqrt{F_{\ell,\ell}^{(1)}F_{\ell,\ell}^{(2)}}}\right)\notag\\
			&-\left(\!\sum_{\ell\neq k}\frac{\frac{\partial |F_{k,\ell}^{(1)}|^{2}}{\partial \widetilde{\phi}_{n}}F_{\ell,\ell}^{(1)}\!-\!\frac{\partial F_{\ell,\ell}^{(1)}}{\partial \widetilde{\phi}_{n}}|F_{k,\ell}^{(1)}|^{2}}{(F_{\ell,\ell}^{(1)})^{2}}\!+\!\frac{\frac{\partial |F_{k,\ell}^{(2)}|^{2}}{\partial \widetilde{\phi}_{n}}F_{\ell,\ell}^{(2)}\!-\!\frac{\partial F_{\ell,\ell}^{(2)}}{\partial \widetilde{\phi}_{n}}|F_{k,\ell}^{(2)}|^{2}}{(F_{\ell,\ell}^{(2)})^{2}} \right.\notag\\
			&\left. +\!2\left[\frac{\partial\Re\left\{F_{k,\ell}^{(1)}F_{k,\ell}^{*(2)}\right\}}{\partial\widetilde{\phi}_{n}}\frac{1}{\sqrt{F_{\ell,\ell}^{(1)}F_{\ell,\ell}^{(2)}}}\!-\!\frac{\partial\sqrt{F_{\ell,\ell}^{(1)}F_{\ell,\ell}^{(2)}}}{\partial\widetilde{\phi}_{n}}\frac{\Re\left\{F_{k,\ell}^{(1)}F_{k,\ell}^{*(2)}\right\}}{F_{\ell,\ell}^{(1)}F_{\ell,\ell}^{(2)}}\right]\right)\notag\\
			&\times\left(F_{k,k}^{(1)}+F_{k,k}^{(2)}+2\sqrt{F_{k,k}^{(1)}F_{k,k}^{(2)}}\right)\Bigg\}\frac{1}{\left(\ds\sigma_{z}^{2}\!+\!\sum_{\ell\neq k}\frac{|F_{k,\ell}^{(1)}|^{2}}{F_{\ell,\ell}^{(1)}}\!+\!\frac{|F_{k,\ell}^{(2)}|^{2}}{F_{\ell,\ell}^{(2)}}\!+\!\frac{2\Re\left\{F_{k,\ell}^{(1)}F_{k,\ell}^{*(2)}\right\}}{\sqrt{F_{\ell,\ell}^{(1)}F_{\ell,\ell}^{(2)}}}\right)^{2}}\notag
		\end{align}
	\end{figure*}
	wherein, for $i\in\{1,2\}$,
	\begin{align}
		\frac{\partial |F_{k,\ell}^{(i)}|^{2}}{\partial \widetilde{\phi}_{n}}&=2\Re\left\{\frac{\partial F_{k,\ell}^{(i)}}{\partial \widetilde{\phi}_{n}}F_{k.\ell}^{(i)*}\right\} \label{partial_derivative_F_k_ell_2} \\ 
		\frac{\partial F_{k,\ell}^{(i)}}{\partial \widetilde{\phi}_{n}}&=\eta_{\ell}^{DL}\rho j\Bigg[\rho \sum_{m\neq n}\left[\mathbf{\widehat{D}}_{i,k}^{H}\mathbf{\widehat{D}}_{i,\ell}\right]_{(m,n)}e^{j(\widetilde{\phi}_{n}-\widetilde{\phi}_{m})} \notag\\
		&-\rho \sum_{i\neq n}\left[\mathbf{\widehat{D}}_{i,k}^{H}\mathbf{\widehat{D}}_{i,\ell}\right]_{(n,i)}e^{-j(\widetilde{\phi}_{n}-\widetilde{\phi}_{i})}\notag\\
		&+e^{j\widetilde{\phi}_{n}}\left[\mathbf{\widehat{D}}_{i,\ell}^{T}\mathbf{\widehat{h}}_{i,k}^{(d)*}\right]_{(n)}-e^{-j\widetilde{\phi}_{n}}\left[\mathbf{\widehat{D}}_{i,k}^{H}\mathbf{\widehat{h}}_{i,\ell}^{(d)}\right]_{(n)} \Bigg] \label{partial_derivative_F_k_ell} \\ 
		&\hspace{-1cm}\frac{\partial\Re\left\{F_{k,\ell}^{(1)}F_{k,\ell}^{*(2)}\right\}}{\partial\widetilde{\phi}_{n}}=\Re\left\{\frac{\partial F_{k,\ell}^{(1)}}{\partial \widetilde{\phi}_{n}}F_{k,\ell}^{*(2)}+\frac{\partial F_{k,\ell}^{*(2)}}{\partial \widetilde{\phi}_{n}}F_{k,\ell}^{(1)}\right\}\label{partial_derivative_F_k_ell_product}\\ 
		&\hspace{-1cm}\frac{\partial\sqrt{F_{k,k}^{(1)}F_{k,k}^{(2)}}}{\partial\widetilde{\phi}_{n}}=\frac{\frac{\partial F_{k,k}^{(1)}}{\partial\widetilde{\phi}_{n}}F_{k,k}^{(2)}+\frac{\partial F_{k,k}^{(2)}}{\partial\widetilde{\phi}_{n}}F_{k,k}^{(1)}}{2\sqrt{F_{k,k}^{(1)}F_{k,k}^{(2)}}} \label{partial_derivative_F_k_ell_product_sqrt}
	\end{align}
	Equipped with the above derivatives, the gradient algorithm can be implemented by means of any off-the-shelf software routine. 
	
	{
		\subsubsection*{Computational complexity}
		The gradient algorithm iteratively updates the variable according to the update rule:
		\begin{equation}
			\bm{\widetilde{\phi}}^{(s)}=\bm{\widetilde{\phi}}^{(s-1)}-\nu \nabla G(\bm{\widetilde{\phi}}^{(s-1)})\;,
		\end{equation}
		wherein $\bm{\widetilde{\phi}}^{(s)}$ denotes the vector of RIS phases at the $s$-th iteration, $\bm{\widetilde{\phi}}^{(s-1)}$ denotes the vector of RIS phases at the $(s-1)$-th iteration, $G$ is the objective function to maximize, and $\nabla G(\bm{\widetilde{\phi}}^{(s-1)})$ is the gradient of $G$ evaluated at $\bm{\widetilde{\phi}}^{(s-1)}$. Finally, $\nu$ is the step-size that defines the magnitude of each update. 
		For the case at hand, before the gradient method starts, it is possible to compute the matrices $\mathbf{\widehat{D}}_{i,k}$, the vectors $\mathbf{\widehat{h}}_{i,k}^{(d)}$, and all the operations involving only these quantities that appear in Eqs. \eqref{F_k_ell_def}-\eqref{partial_derivative_F_k_ell_product_sqrt}. Thus, the complexity associated with these operations will be neglected in the sequel, since these are simple initializations that need not be repeated in each iteration of the gradient method.
		Instead, the bulk of the complexity is due to the computation of $\bm{\widetilde{\phi}}^{(s)}$ in each iteration, times the number of iterations until convergence. The former can be computed by the formulas in Eq. \eqref{Eq:DerG}, which, although cumbersome to write, contains only elementary functions and thus can be easily evaluated based on the formulas in Eqs. \eqref{F_k_ell_def} and \eqref{partial_derivative_F_k_ell}. Inspecting \eqref{F_k_ell_def}, we can see that its complexity scales as $O(N_{R}N_{B})$, because $N_{R}N_{B}$ multiplications are required to compute each of the products ${\mathbf{\widehat{D}}_{i,k}}e^{j\bm{\widetilde{\phi}}}$ and ${\mathbf{\widehat{h}}_{i,\ell}^{(d)}}e^{j\bm{\widetilde{\phi}}}$. Then, \eqref{F_k_ell_def} must be computed $K^{2}$ times, i.e. for any $k$ and $\ell$. As for \eqref{partial_derivative_F_k_ell}, its complexity is linear in $N_{R}$ due to the sum over the number of RIS elements, and again quadratic in $K$, since it must be computed for any $k$ and $\ell$. Once \eqref{F_k_ell_def} and \eqref{partial_derivative_F_k_ell} have been computed, \eqref{partial_derivative_F_k_ell_2} , \eqref{partial_derivative_F_k_ell_product}, and \eqref{partial_derivative_F_k_ell_product_sqrt} can be computed by one, two, or three additional multiplications for each $n=1,\ldots,N$ and $k,\ell=1,\ldots,K$. Finally, all these quantities can be plugged into \eqref{Eq:DerG}, which requires an additional number of computations that is quadratic in $K$, due to the nested sums over the number of users. Thus, asymptotically, the complexity of computing ${\bm \phi}^{(s)}$ in each iteration scales as the complexity of computing \eqref{F_k_ell_def}, which is $O(K^{2}N_{R}N_{B})$, and the overall asymptotic complexity of the considered instance of the gradient method is $O(N_{it}K^{2}N_{R}N_{B})$, where $N_{it}$ is the number of iterations until convergence. 
		As for $N_{it}$, it is challenging to give a closed-form expression, as it heavily depends on the choice of the step-size $\nu$. Moreover, $\nu$ is typically handled in an adaptive way, i.e. updating its value during the execution of the algorithm, in order to achieve the best trade-off between convergence speed and performance. Empirically, in our simulations we have observed convergence in a handful of iterations.}
	
	\subsection{Solution of the problem with respect to $\bm{\eta}_1^{\rm DL}$ and $\bm{\eta}_2^{\rm DL}$} \label{Eta_Opt_Section}
	After the solution of Problem \eqref{Prob:MaxSNR_MU1_phi}, we solve now the problem with respect to the variables $\bm{\eta}_1^{\rm DL}$ and $\bm{\eta}_2^{\rm DL}$ for fixed $\bm{\widetilde{\phi}}$. To begin with, we again take the logarithm of the objective function of \eqref{Prob:MaxSNR_MU2}, which causes no optimality loss, since the logarithm is an increasing function. Thus, the optimization problem with respect to the transmit powers can be equivalently reformulated as 
	
	\begin{subequations}\label{Prob:MaxSNR_MU2_eta}
		\begin{align}
			\ds\max_{\bm{\eta}^{\rm DL}} \sum_{k=1}^K  \; & \ds  \log_2 \left(\frac{\left|\ds \sum_{i=1}^2 I_{i,k}\sqrt{\eta_{i,k}^{\rm DL}} a_{k,k}^{(i)}\right|^2}{\ds \sum_{\substack{\ell=1 \\ \ell \neq k}}^K {\left|\ds \sum_{i=1}^2 I_{i,\ell}\sqrt{\eta_{i,\ell}^{\rm DL}} a_{k,\ell}^{(i)}\right|^2}  +  \sigma^2_z}\right) \, , \\
			\textrm{s.t.} \quad & \sum_{\ell=1}^K {I_{i,\ell}\eta_{i,\ell}^{\rm DL}} \leq P_{\rm max}^{{\rm BS},i}, \; , i=1,2\\
			&  {\eta_{i,\ell}^{\rm DL}} \geq 0 \; \forall \ell=1,\ldots,K \; , i=1,2.,  
		\end{align}
	\end{subequations} 
	with
	\begin{equation}
		a_{k,\ell}^{(i)}= \ds \frac{\left( \rho \widehat{\bD}_{i,k} e^{j\bm{\widetilde{\phi}}} +\widehat{\bh}_{i,k}^{(d)}\right)^{H}\left( \rho \widehat{\bD}_{i,\ell} e^{j\bm{\widetilde{\phi}}}  +\widehat{\bh}_{i,\ell}^{(d)} \right)}{\norm{\rho \widehat{\bD}_{i,\ell} e^{j\bm{\widetilde{\phi}}}  +\widehat{\bh}_{i,\ell}^{(d)}}} \, .
	\end{equation}
	Problem \eqref{Prob:MaxSNR_MU2_eta} can be tackled by means of the sequential optimization framework. To see this, let us expand the objective of \eqref{Prob:MaxSNR_MU2_eta} as the difference of two logarithms, which yields
	\begin{align}\label{Eq:ObjPower}
		\text{SR}&=\sum_{k=1}^{K}\log_{2}\left(I_{1,k} \eta_{1,k}^{DL}(a_{k,k}^{(1)})^{2}+I_{2,k} \eta_{2,k}^{DL}(a_{k,k}^{(2)})^{2}\right.\notag\\
		&\hspace{3cm}\left.+2 I_{1,k}I_{2,k}\sqrt{\eta_{1,k}^{DL}\eta_{2,k}^{DL}}a_{k,k}^{(1)}a_{k,k}^{(2)}\right)\notag\\
		&-\sum_{k=1}^{K}\log_{2}\Bigg(\sigma_{z}^{2}+\sum_{\ell\neq k}I_{1,\ell}\eta_{1,\ell}^{DL}(a_{k,\ell}^{(1)})^{2}+I_{2,\ell}\eta_{2,\ell}^{DL}(a_{k,\ell}^{(2)})^{2}\notag\\
		&\hspace{3cm}+2I_{1,\ell}I_{2,\ell}\sqrt{\eta_{1,\ell}^{DL}\eta_{2,\ell}^{DL}}a_{k,\ell}^{(1)}a_{k,\ell}^{(2)}\Bigg)
	\end{align}
	By virtue of \cite[Lemma 1]{Dandrea2020} \eqref{Eq:ObjPower} is the difference of two concave functions. Thus, it can be maximized by an instance of the sequential optimization method, in which at the $j$-th iteration the second line in \eqref{Eq:ObjPower} is linearized around the point $(\bm{\tilde{\eta}}_{1},\bm{\tilde{\eta}}_{2})$ obtained as the solution of the linearized problem at the $(j-1)$-th iteration.
	
	\begin{table}
		\centering
		\caption{Simulation parameters}
		\label{table:parameters}
		\def\arraystretch{1.2}
		\begin{tabulary}{\columnwidth}{ |p{3cm}|p{4.5cm}| }
			\hline
			MSs distribution 				& Horizontal: uniform in each cell, vertical: 1.5~m\\ \hline
			BS height				& 25~m\\ \hline
			RIS height				& 40~m\\ \hline
			Carrier freq., bandwidth		&  $f_0=3$ GHz, $B = 20$ MHz \\ \hline
			BSs antenna array			& 16-element with $\lambda/2$ spacing\\ \hline
			RIS antenna array			& $N_R$-element with $\lambda/2$ spacing\\ \hline
			MS antennas 		& Omnidirectional with 0~dBi gain\\ \hline
			Thermal noise 				& -174 dBm/Hz spectral density \\ \hline
			Noise figure 			& 9 dB at BS/MS\\ \hline
		\end{tabulary}
	\end{table}
	
	\section{Numerical Results}
	In order to provide numerical results, we refer to the scenario depicted in Fig. \ref{Fig:Scenario_multi_BS}, considering an inter-site distance between the two BSs of 300 meters. We consider BSs with 16 antennas, i.e., $N_{B,1} =N_{B,2}=64$ and a RIS with $N_R$ reflecting elements; the number of MSs in each cell is 10, thus the number of users simultaneously served on the same frequency in the system is $K=20$ in the multi-user scenario. All the remaining simulation parameters are summarized in Table \ref{table:parameters}.
	
	The channel coefficients $\beta_{i,k}$ and $\beta_{i,k}^{(d)}$ that model the attenuation on the reflected path $k$-th MS-RIS-BS and on the direct $k$-th MS-BS path with respect to the $i$-th BS, respectively, are modelled according to \cite{EE_RISs}. In particular,
	\begin{equation}
		\beta_{i,k}=\frac{10^{-3.53}}{\left(d_{{\rm BS}_i ,\rm{RIS}}+d_{{\rm RIS},k}\right)^{3.76}} \, , \; \; \text{and} \; \; \beta_{i,k}^{(d)}=\frac{10^{-3.53}}{d_{{\rm BS}_i,k}^{3.76}},
		\label{Beta_k}
	\end{equation}
	where $d_{{\rm BS}_i ,\rm{RIS}}$ and $d_{{\rm RIS},k}$ are the distance between the $i$-th BS and the RIS and the distance between the $k$-th MS and the RIS in meters, respectively, while $d_{{\rm BS}_i,k}$ is the distance between the $i$-th BS and the $k$-th MS. 
	
	We assume that the maximum power transmitted by the BS is $P_{\rm max}^{{\rm BS},i}=10$ W. In the following figures, we report the results of the resource allocation strategies proposed in the paper in the case of   {perfect channel state information (PCSI)} and in the case in which channel estimation (CE) is performed at each BS according to the procedures detailed in Section \ref{Section_CE}. With regard to the CE procedure, orthogonal sequences with length $\tau_p=K$ are assumed at the MS and the power transmitted during the uplink training is $\eta_k=\tau_p\widetilde{\eta}_k$ with $\widetilde{\eta}_k=100$ mW. The number of RIS configurations during the CE procedures is $Q=N_R+1$ in the case of LS CE and MMSE with $Q$ realizations of the RIS (MMSEQ), and $Q=1$ in the case of MMSE with one random realization of the RIS configuration (MMSE1).
	
	\begin{figure}[!t]
		\centering
		\includegraphics[scale=0.63]{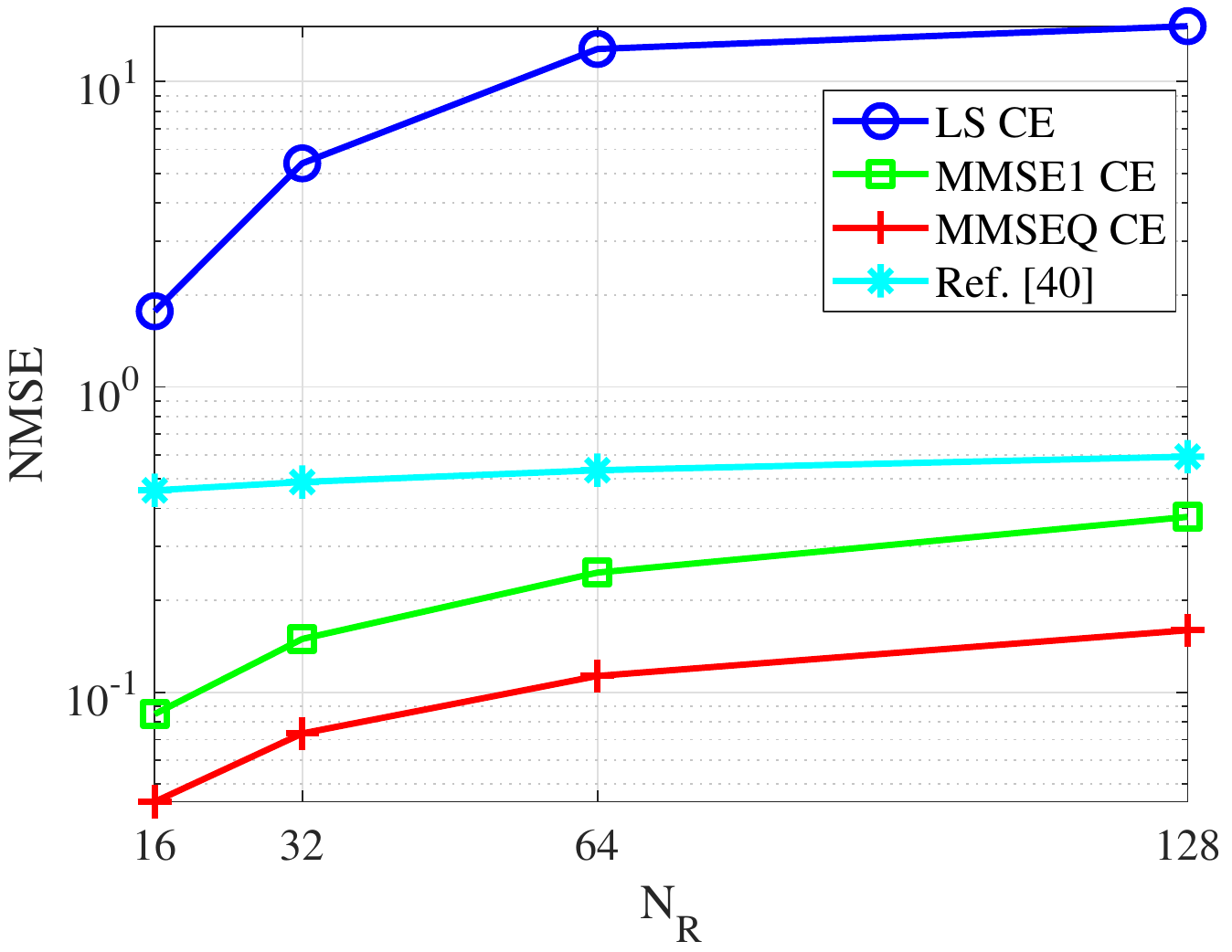}
		\caption{NMSE versus $N_R$ in the multiuser scenario with $N_B=64$ and $K=20$.}
		\label{Fig:NMSE_comparison}
	\end{figure}
	
	{
		\subsection{Channel estimation performance}
		In this subsection, we evaluate the performance of the channel estimation procedures proposed in the paper and provide a comparison with the channel estimation framework proposed in reference \cite{Wang2019d}. The channel estimation procedure proposed in \cite{Wang2019d} is based on the idea of switching groups of RIS elements on and off and assumes a three-phases protocol in which i) the RIS is switched off and the BS estimates the direct channel between BS and MS, ii) RIS is switched on and only one user transmits pilot signals, and iii) groups of elements are switched on and all the other users transmit pilot signals alternatively. In reference \cite{Wang2019d} the authors also use an MMSE approach and assume knowledge of the statistics of the correlation between channels to different users. The pilot training required by the channel estimation in \cite{Wang2019d} is lower, in fact the minimum number of pilots is $K+N_R+\max\left( K-1, \lceil \frac{(K-1)N_R}{N_B} \rceil  \right)$, while our approach requires $K(N_R+1)$ total pilot signals for LS CE and MMSEQ CE and $K$ in MMSE CE. 
		We report the performance of the considered channel estimation procedures in terms of normalized mean-squares-error (NMSE) defined as in Eq. \eqref{NMSE_definition} at the top of next page.
		\begin{figure*}
			{\begin{equation}
					\text{NMSE}_i=\frac{\ds \sum_{k=1}^K \mathbb{E} \left[ \norm{\bh_{i,k}^{(d)}-\widehat{\bh}_{i,k}^{(d)}}^2 \right] +  \ds \sum_{n=1}^{N_R} \ds \sum_{k=1}^K \mathbb{E} \left[ \norm{\left[\bD_{p,k}\right]_{(:,n)}-\left[\widehat{\bD}_{p,k}\right]_{(:,n)}}^2 \right]}{\ds \sum_{k=1}^K \mathbb{E} \left[ \norm{\bh_{i,k}^{(d)}}^2\right] +  \ds \sum_{n=1}^{N_R} \ds \sum_{k=1}^K \mathbb{E} \left[ \norm{\left[\bD_{p,k}\right]_{(:,n)}}^2 \right]}
					\label{NMSE_definition}
			\end{equation}}
		\end{figure*}
		In Fig. \ref{Fig:NMSE_comparison}, we report the performance of the LS CE, MMSE1 CE and MMSEQ CE in terms of NMSE versus the number of antennas at the RIS $N_R$. We can note that the MMSE-based approaches offer better performance with respect to the LS one, that does not assume knowledge of the large scale coefficients of the channels. Assuming the same amount of channel knowledge and length of the pilot signals, we report also the performance of the procedure proposed in reference \cite{Wang2019d} and we note that with MMSE approaches we outperform it. 
		Our approaches thus exhibit the following advantages. First, better performance in terms of NMSE is obtained in the considered scenario, and, then, it is not required switching on/off the RIS elements and making them completely absorbing, which may be difficult.}
	
	\subsection{Single-user and single BS}
\begin{figure*}[!t]
		\centering
		\includegraphics[scale=0.45]{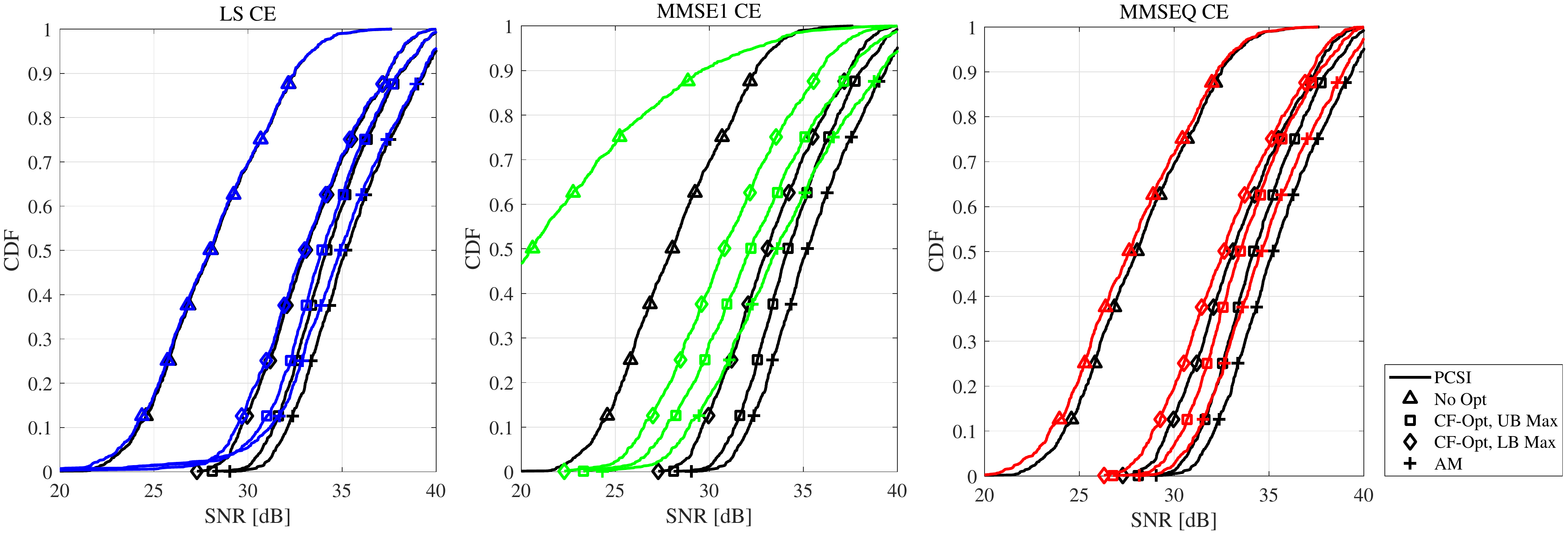}
		\caption{CDFs of the SNR in the single-user scenario with $N_R=64$ and $N_B=64$. Performance obtained in the following cases: closed-form optimization via the upper-bound maximization (CF-Opt, UB Max); closed-form optimization via the lower-bound maximization (CF-Opt, LB Max); alternating maximization (AM) approach and random configuration of the RIS (No Opt.).}
		\label{Fig:SNR_SingleUser_Opt}
	\end{figure*}

	We start evaluating the performance of the optimization procedure detailed in Section \ref{Single_user_Resource}.
	In Fig. \ref{Fig:SNR_SingleUser_Opt}, we report the cumulative distribution functions (CDFs) of the SNR. In particular, we compare the performance obtained in the following cases: closed-form optimization (CF-Opt) via the upper-bound maximization (UB Max) in Section \ref{UB_max_Section}; closed-form optimization (CF-Opt) via lower-bound maximization (LB Max) in Section \ref{LB_Max_Section}; alternating maximization (AM) approach and random configuration of the RIS (No Opt.).   {In the AM approach we use alternating optimization to find a candidate solution of Problem \eqref{Prob:MaxSNR} \cite[Section 2.7]{BertsekasNonLinear}, i.e, we solve alternatively the problem with respect to $\bm{\phi}$ and then the problem with respect to $\mathbf{w}$.} We can see that the proposed strategies are effective since the gap with the performance corresponding to the case in which the RIS configuration is random is of several dBs.  
	Moreover, the performance of the closed-form solutions is very close (within few dBs) to the one obtained using the AM  methodology, with a significantly lower complexity, given the fact that we have been able to express the solution in closed form. The figure also shows that for the case in which MMSE CE with $Q>1$ RIS configurations is used the performance are quite close (within fractions of dB) to the ones attained in the case of perfect CSI. 
	
\begin{figure}[!t]
		\centering
		\includegraphics[scale=0.55]{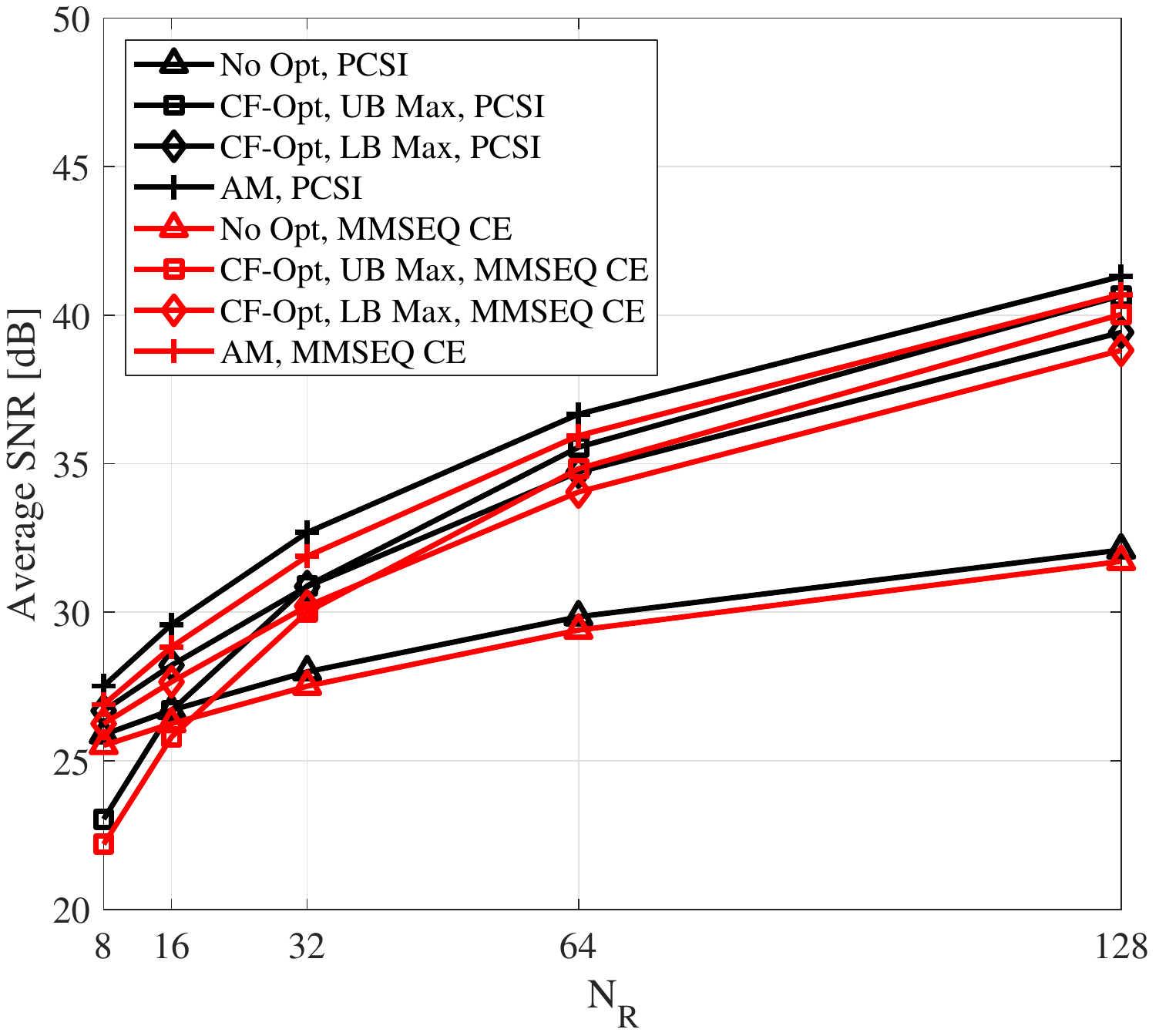}
		\caption{Average SNR versus $N_R$ in the single-user scenario with $N_B=64$. Performance obtained in the following cases: closed-form optimization via the upper-bound maximization (CF-Opt, UB Max); closed-form optimization via the lower-bound maximization (CF-Opt, LB Max); alternating maximization (AM) approach and random configuration of the RIS (No Opt.).}
		\label{Fig:SNR_SingleUser_vsNR}
	\end{figure}

	Fig. \ref{Fig:SNR_SingleUser_vsNR} shows the impact on the system performance of the number of the RIS reflecting elements. The figure shows that, for the considered optimization procedures, when $N_R$ is increased from 8 to 128, the SNR gain is in the order of about 15 dB. The SNR values corresponding to the point $N_R=8$ is also an upper bound to the system performance when no RIS is present in the system. This plot thus also shows that installing a RIS may bring very remarkable performance improvements.

	\subsection{Performance of multi-user system with joint transmission}
	We now evaluate the performance obtained by the optimization procedure described in the general multiuser scenario with two BSs and joint transmission, as reported in Section \ref{Joint_Resource}.
	First of all, we detail the procedure that we use to define the binary variables $I_{i,\ell}, \, \forall i =1,2, \text{and} \, \ell=1,\ldots K$, i.e. to determine which users are to be jointly served by the two BSs in the system. 
	Basically, we choose to adopt joint transmission for a fixed percentage, $p_{\rm JT}$ say, of the users in the system, and denote by $K_{\rm JT}$ the number of MSs served by joint transmission.  More precisely, the procedure starts by associating each MS to the BS with the largest direct power attenuation coefficient; otherwise stated, the generic $k$-th MS is associated to the $i^*$-th BS such that
	\begin{equation}
		i^*= \arg \max_{i=1,2} \beta_{i,k}^{(d)} \, .
	\end{equation}
	Then, the $K_{\rm JT}$ users with the lowest value of the metric 
	\begin{equation}
		\gamma_{k}^{(d)}=\frac{\max \left( \beta_{1,k}^{(d)}, \beta_{2,k}^{(d)}\right)}{\min \left( \beta_{1,k}^{(d)}, \beta_{2,k}^{(d)}\right)} \, 
	\end{equation}
	are served with joint transmission. 
	
\begin{figure*}[!t]
		\centering
		\includegraphics[scale=0.45]{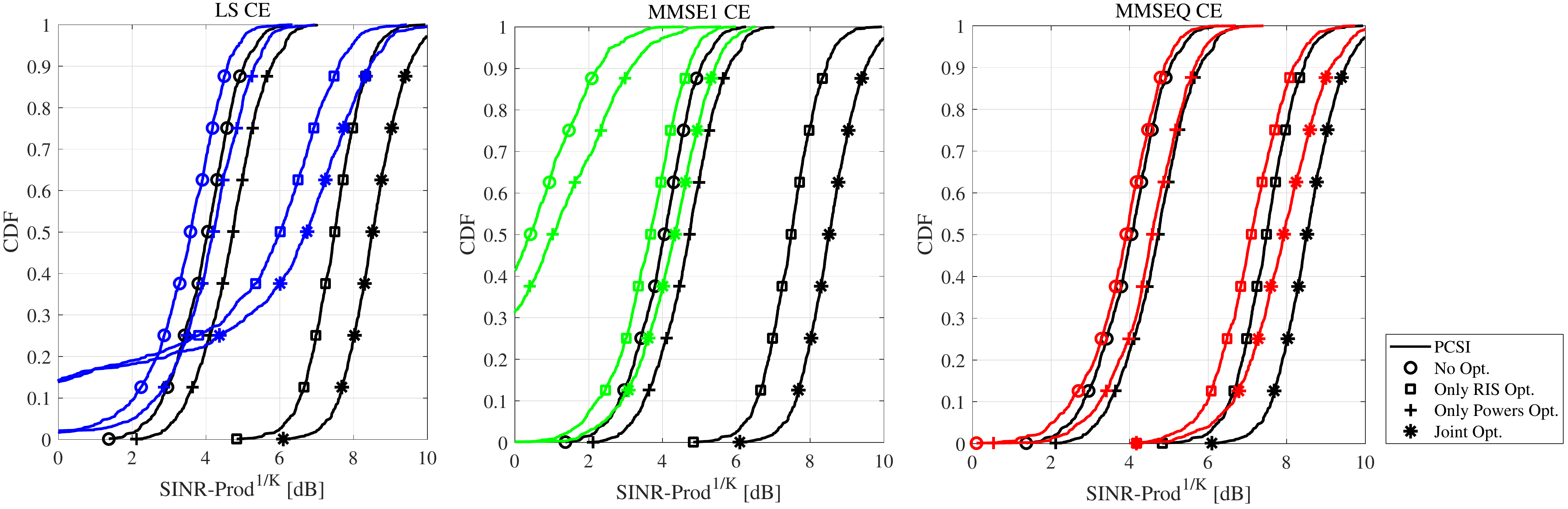}
		\caption{CDFs of the geometric mean of the SINRs in the multi-user scenario with $N_R=64$, $N_{B,1}=N_{B,2}=64$, $K=20$ and $p_{\rm JT}= 20 \%$. Performance obtained in the following cases: joint passive beamforming design and power allocation (Joint Opt.) technique; optimization of the only RIS phase shifts (Only RIS Opt.); optimization of the only transmit powers (Only Powers Opt.) and random configuration of the RIS with uniform power allocation (No Opt.).}
		\label{Fig:SNR_MultiUser_Opt}
	\end{figure*}

	Fig. \ref{Fig:SNR_MultiUser_Opt} reports the CDFs of the geometric mean of the SINRs in the multi-user scenario. In particular, we compare the performance obtained in the following cases:   {joint passive beamforming design and power allocation (Joint Opt.); optimization of the only RIS phase shifts (Only RIS Opt.); optimization of the only transmit powers (Only Powers Opt.), and without optimization (No Opt.). In the Joint Opt., we use the whole procedure proposed in Section \ref{Joint_Resource}; in Only RIS Opt., we optimize the RIS configuration as in Section \ref{Phi_opt_Section} and consider uniform power allocation, in Only Powers Opt., we assume a random configuration of the phase shifts and optimize the transmit power according to the procedure in Section \ref{Eta_Opt_Section}, and in No Opt., we assume random configuration of the RIS and uniform power allocation.} It is assumed that 20\% of the users are served with joint transmission, i.e., $p_{\rm JT}= 20 \%$. The results confirm that the proposed procedures are effective, since noticeable performance gains are obtained with respect to the case in which the RIS has a random configuration. The joint optimization of the RIS phase shifts and of the transmit powers achieves the best performance; also in this case MMSEQ channel estimation exhibits the smallest gap with respect to the ideal performance obtained with perfect CSI. 
	
\begin{figure}[!t]
		\centering
		\includegraphics[scale=0.47]{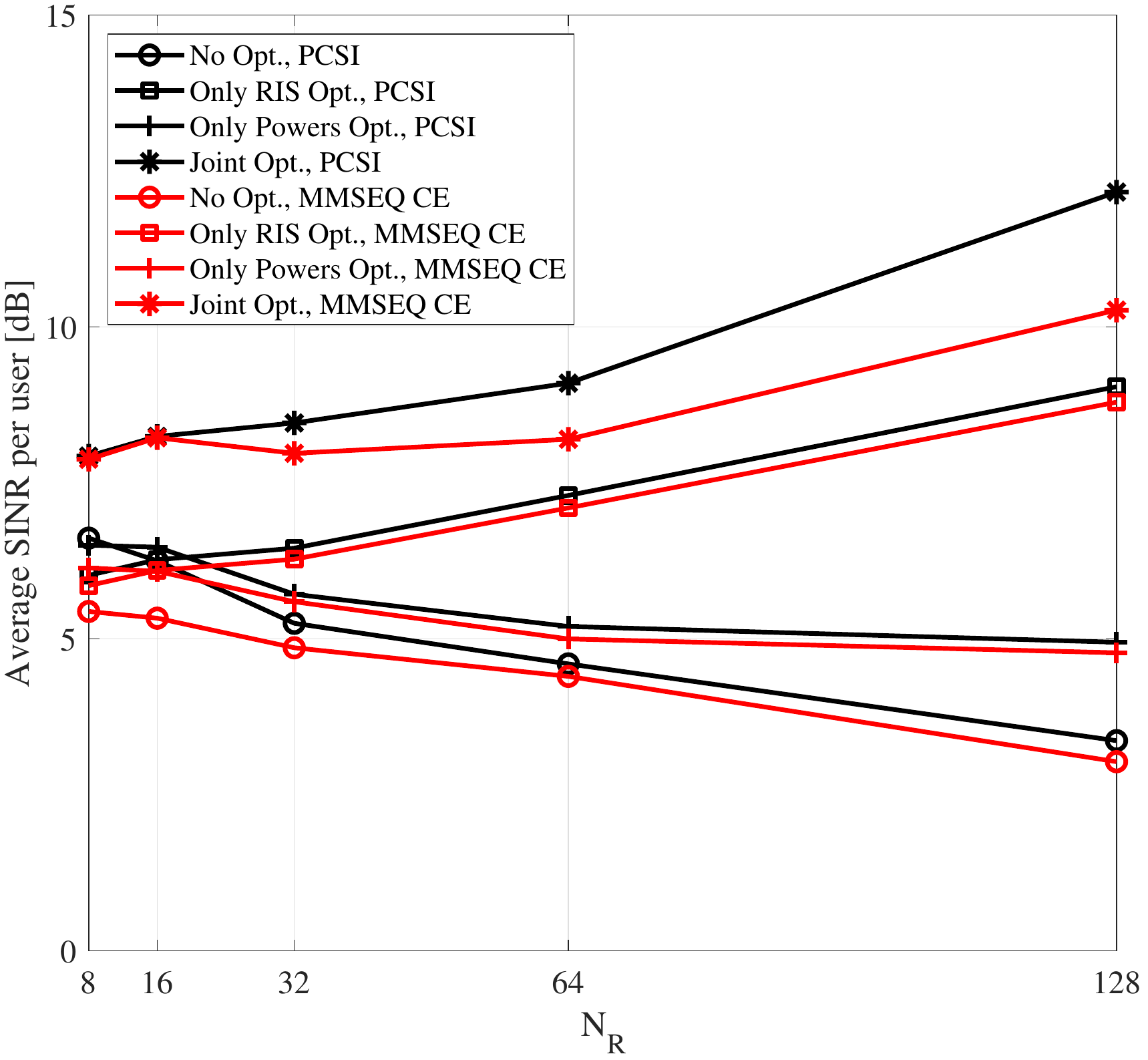}
		\caption{Average SINR versus number of reflecting elements at the RIS, $N_R$, $N_{B,1}=N_{B,2}=64$,$K=20$, $p_{\rm JT}= 20 \%$ and MMSEQ CE. Performance obtained in the following cases: joint passive beamforming design and power allocation (Joint Opt.) technique; optimization of the only RIS phase shifts (Only RIS Opt.); optimization of the only transmit powers (Only Powers Opt.) and random configuration of the RIS with uniform power allocation (No Opt.).}
		\label{Fig:Av_SNR_vs_NR}
	\end{figure}

	Fig. \ref{Fig:Av_SNR_vs_NR} shows the impact of the number of reflecting elements at the RIS, $N_R$, on the the average SINR per user, using the same resource allocation techniques reported in Fig. \ref{Fig:SNR_MultiUser_Opt}. One very interesting remark that can be done here is that in a multiuser environment, when the RIS phase shifts are not optimized, the performance slightly decreases when the RIS size increases. This result can be justified by noticing that a RIS with random phase shifts increases the overall interference level and ultimately decreases the system performance. When, instead, RIS phases are optimized, the system performance correctly increases, even though the improvements are now smaller than those observed in Fig. \ref{Fig:SNR_SingleUser_vsNR} for a single user scenario. This behavior can be explained by noticing that in a single-user system all the RIS elements can be exploited in order to improve the system performance of the only user in the system, while in the present multiuser system the RIS benefit is to be shared among the several users in the system. 
	
\begin{figure}[!t]
		\centering
		\includegraphics[scale=0.47]{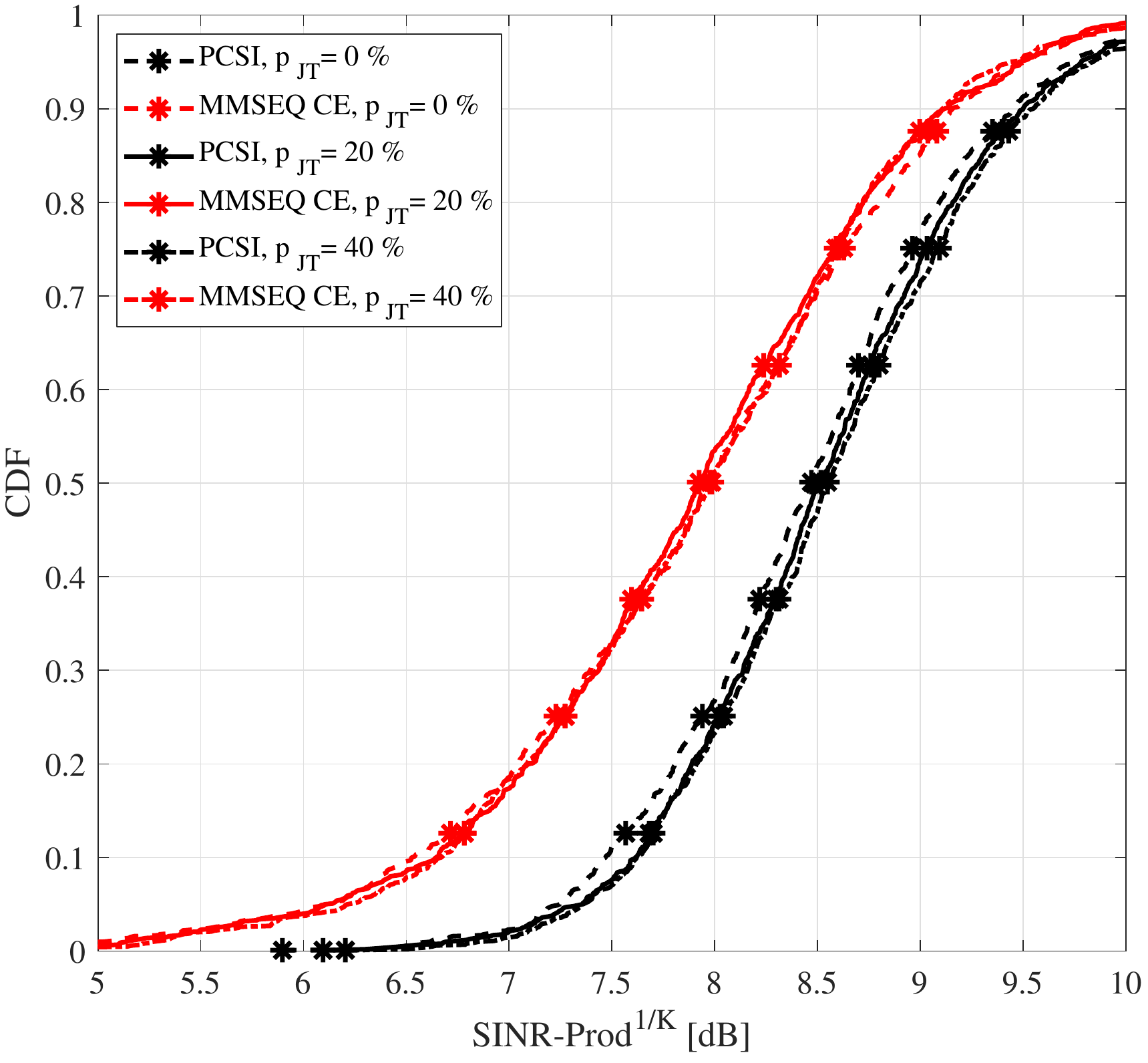}
		\caption{CDFs of the geometric mean of the SINRs in the multi-user scenario with different values of $p_{\rm JT}$, $N_R=64$, $N_{B,1}=N_{B,2}=64$, $K=20$, MMSEQ CE and with the joint passive beamforming design and power allocation (Joint Opt.)}
		\label{Fig:CDF_SINR_geom_mean_CoMP}
	\end{figure}

	Finally, Fig. \ref{Fig:CDF_SINR_geom_mean_CoMP} shows the CDF of the geometric mean of the SINR for several values of the parameter $p_{\rm JT}$. The general behaviour that is observed is that usually increasing the number of users enjoying joint transmission brings a performance improvement for most of the users in the system, even though a few fraction of them (the ones represented in the highest part of the CDF) experience some performance degradation due to the increased level of interference. These ones are indeed the users very close to the serving BS, and the activation of the joint transmission to cell-edge users causes a reduction of their received useful power. 
	
	Overall, the plots show that the proposed resource allocation procedures bring considerable performance improvements to the system, that the presence of the RIS is beneficial to the system, and, also, that the proposed CE techniques blend well with the described optimization procedures.

	\section{Conclusion}
	For a wireless network assisted by a RIS, channel estimation algorithms and resource allocation procedures have been presented in this paper. In particular, the paper has tackled the problem of SNR maximization with respect to  the RIS configuration and to the BS beamformer for a single-user setting. Moreover, for a multi-user multi-cell scenario, the geometric mean of the SINRs has been maximized with respect to the BS transmit power vectors and to the RIS configuration, assuming that some of the users are jointly served by two BSs. The obtained results have shown the beneficial impact on the system performance of the presence of a RIS and of the described optimization procedures. Current research on this topic is now centred on the study of RISs in conjunction with innovative networks deployments such as cell-free massive MIMO systems.

	\ifCLASSOPTIONcaptionsoff
	\newpage
	\fi
	
	\bibliographystyle{IEEEtran}
	\bibliography{LISreferences,FracProg,RIS_Survey}

\end{document}